# Improving full-waveform inversion based on sparse regularisation for geophysical data

**Jiahang Li**[1*], **Hitoshi Mikada**[1,2] **and Junichi Takekawa**[1]

[1]*Department of Civil and Earth Resources Engineering, Kyoto University, Kyoto, Japan*

[2]*Centre for Integrated Research and Education of Natural Hazards, Shizuoka University, Shizuoka Prefecture, Japan*

**Corresponding author**: Jiahang Li. E-mail: jiahang.li@tansa.kumst.kyoto-u.ac.jp

**Abstract**

Full Waveform Inversion (FWI) is an advanced geophysical inversion technique. In fields such as oil exploration and geology, FWI is used for providing images of subsurface structures with higher resolution. The conventional algorithm minimises the misfit error by calculating the least squares of the wavefield solutions between observed data and simulated data, followed by gradient direction and model update increment. Since the gradient is calculated by forward and backward wavefields, the high-accuracy model update relies on accurate forward and backward wavefield modelling. However, the quality of wavefield solutions obtained in practical situations could be poor and does not meet the requirements of high-resolution FWI. Specifically, the low-frequency wavefield is easily affected by noise and downsampling, which influences data quality, while the high-frequency wavefield is susceptible to spatial aliasing effects that produce imaging artefacts. Therefore, we propose using an algorithm called Sparse Relaxation Regularised Regression to optimise the wavefield solution in frequency domain FWI, which is the forward and backward wavefield obtained from the Helmholtz equation, and thus improve the accuracy of the FWI. The sparse relaxed regularised regression algorithm combines sparsity and regularisation, allowing the broadband FWI to reduce the effects of noise and outliers, which can provide data supplementation in the low-frequency band and anti-aliasing in the high-frequency band. Our numerical examples demonstrate the wavefield optimization effect of the sparse relaxed regularised regression-based algorithm in various cases. The accuracy and stability of the improved algorithm are verified in comparison to the Tikhonov regularisation algorithm.

# 1. Introduction

Full waveform inversion (FWI) is a powerful technique for constructing high-resolution models of subsurface seismic velocity structures, which is extensively employed in geophysical exploration, seismic monitoring, and oil and gas reservoir characterization (Virieux & Operto 2009). In the frequency domain FWI, the seismic modelling technique simulates the forward and backward wavefield solution by solving the wave equation. Subsequently, simulated and observed seismic records are matched iteratively to solve the inverse problem and obtain a velocity model (Warner & Guasch 2016).

However, FWI encounters several challenges, such as tending to converge to local minima, azimuth and acquisition system limitations, and imaging artefacts. Especially when there is numerical error in the forward data and low-density acquisition systems, both will greatly affect the inversion results (Zhang 2010). The frequency-domain forward modelling generates a forward and backward wavefield that depicts the seismic wavefield propagation relationship between the source and receiver at a specific frequency (Wu & McMechan, 2021). Especially, the three-dimensional wavefield stack contains multi-scale information, ranging from low to high frequencies, which can be utilised to represent the structures and properties of the subsurface at varying scales (Ovcharenko *et al.* 2018). However, in real cases, the forward and backward wavefield is often influenced by numerical error, leading to not accurate forward wavefields and further resulting in degraded quality of imaging and inversion outcomes (Alkhalifah *et al.* 2021). Although many scholars try to develop new schemes for accurate seismic modelling in the frequency domain (Jo *et al.* 1996; Chen 2012; Takekawa & Mikada 2018), the increment of the computational costs is generally inevitable, especially in field studies. It makes it difficult to apply FWI in practice. More importantly, in situations where high-density acquisition is challenging to attain, the effect of spatial aliasing must be considered. The spatial aliasing effect arises from the unsatisfied Nyquist-Shannon sampling theorem during the sampling process, which leads to the possibility of high-frequency components being misinterpreted as low-frequency components. It causes low-frequency information to be lost, consequently impacting the accurate recovery of the subsurface velocity structure (Shaiban *et al.* 2022). The most intuitive effect of this phenomenon is that there will be high-frequency artefacts in the high-frequency band. The traditional approach of increasing the sampling rate will augment the amount of data but may also elevate the computational cost (Liu & Fomel 2011).

Furthermore, FWI encounters numerous issues in cases of low-density sampling (which implies that the seismic data are sparsely sampled). Firstly, it is well known that low-frequency information is essential for FWI, as it helps to overcome the nonlinearity problem, thereby facilitating the convergence of the inversion process with abundant low-frequency information to the global optimal solution rather than the local optimal solution (Kan *et al.* 2023). Additionally, low-density sampling may contribute to the limited offset distance issue, that is, insufficient receiver coverage and a decline in the inversion resolution of the subsurface structure (Operto *et al*. 2009). These limitations can render the FWI update process more reliant on the initial model, diminish the accuracy of the inversion results, and impact the identification and interpretation of subsurface structures by FWI (Operto & Miniussi 2018).

Based on the above problems, we propose to use the sparse relaxation regularised regression (SR3) algorithm in FWI to optimise the forward and backward wavefield obtained from the frequency-domain numerical modelling to improve the accuracy of FWI and achieve high-resolution inversion (Zheng *et al.* 2019). The SR3 algorithm, as an optimization algorithm, enhances the basic regression problem into a new regression problem with additional constraints through the assistance of auxiliary matrices. The main difference between the SR3 algorithm and conventional regression problems is that it tightens the function space of the error function, which can compress the nonlinear wavefield loss space, thereby shrinking the range of singular values. As a result, the iteration speed of regression based on the SR3 algorithm is faster than that of conventional regression algorithms, and it also encourages sparse solutions (Champion *et al.* 2020). Specifically, a more appropriate image processing technique for the spatial aliasing effect that may occur in high-frequency data is to use matrix reconstruction and interpolation techniques (Liu & Fomel 2011). The SR3, as a regularisation algorithm, has a certain interpolation effect by introducing regularisation terms and sparse constraints. The SR3 can reduce the effect of aliasing, while sparse constraints help to reduce the complexity of the solution space, thus improving the accuracy of matrix reconstruction and interpolation (Brunton *et al.* 2016). In addition, the processing of low-frequency data, as well as data completion and interpolation of down-sampled data, are also crucial for FWI. The SR3 algorithm adopts multi-regularisation algorithms through regression and iterative processes to regularise the sparse basis of the simulated data, and through fast iterations, the data is recovered and completed in the solution of the optimization problem (Erichson *et al.* 2020). Moreover, this algorithm can find a balance between the sparse constraints and the smoothness of the solution by choosing an appropriate regularisation parameter, thereby providing a more stable denoising and data completion effect (Champion *et al.* 2020).

In this paper, based on the frequency-domain FWI, the SR3 algorithm is innovatively introduced to achieve data complementation and reconstruction of FWI for low-density acquisition data, anti-spatial blending effect for high-frequency data, and improved resolution for high background noise datasets. Two homogeneous medium models and two synthesis models (2004 BP Model and SEG/EAGE Overthrust Model) are used to verify the superiority of our approach by comparing Tikhonov regularised FWI with SR3-based FWI.

## 2. Sparse relaxed regularised regression

In the field of FWI technology, the frequency-domain seismic modelling technique has garnered considerable attention and research due to its outstanding precision, efficient computational speed, and independence from overly idealised time sampling rates (Jiang *et al.* 2021). The frequency-domain seismic modelling technique does not rely on an excessively idealised time sampling rate, thus offering greater flexibility in practical applications. A crucial part of this algorithm typically involves solving the Helmholtz equation (Appelö *et al.* 2020). This foundational equation $\mathbf{U}(\mathbf{X},\mathbf{Z};\mathbf{W})$ provides an essential mathematical framework for our understanding of geophysical phenomena:

$$(\Delta + \mathbf{W}^2 \mathbf{M})\mathbf{U} = -F(\mathbf{W})G(\mathbf{X}-\mathbf{X}_S)g(\mathbf{Z}-\mathbf{Z}_S), \tag{1}$$

where $\mathbf{W}$ is the angular frequency, $\mathbf{M}$ is the squared slowness, $F(\mathbf{W})$ is the source signature, $\Delta$ is the Laplacian operator, and $(\mathbf{Z}-\mathbf{Z}_S)$ is the source coordinates.

In the subsequent steps, the conventional algorithm ingeniously maps the model space to the data space, enabling a better understanding and handling of the model. Within the data space, FWI performs a least squares calculation on the discrepancy between observed data and calculated data to assess the agreement between them (Alkhalifah *et al.*, 2021). The resultant least squares value reflects the model increment in the model space, representing the extent of improvement from the current state of the model to a more optimal state. FWI then utilises this model increment to compute the update gradient, which not only expresses the direction of model change but also reflects the rate of model change, providing a guide for optimizing the model (Choi & Alkhalifah 2012). Equipped with this gradient, we are well-prepared for the next iteration: exploring a more optimal model based on the current model and gradient, enhancing the agreement between observed data and calculated data, and thereby attaining a more precise and reliable model:

$$\mathbf{M}_i = \arg \min_{\mathbf{M} \in \mathbf{M}_A} \|F(\mathbf{M}) - \mathbf{D}_{obs}\|, \tag{2}$$

where $F(\mathbf{M})$ is the calculated data and $\mathbf{D}_{obs}$ is the observed data.

However, in the minimization process, the effectiveness and accuracy of the inversion results are directly impacted if the quality of the calculated data matrix we are using is low or if data loss has occurred due to low-density sampling, even when more advanced inversion algorithms are applied. The integrity and quality of data play a crucial role in this process, with data loss and noise causing the inversion results to deviate from true values (Liu & Fomel 2011). Therefore, we employ an optimization algorithm to complete data and denoise the calculated data matrix before the minimization operation, thereby enhancing the quality and completeness of the data and consequently improving the accuracy of inversion results.

In industrial and scientific realms, regularisation regression problems are prevalent; sparse regression plays a vital role in applications such as compressive sensing and data completion. Sparse regression allows us to select the most influential features from large datasets containing many unimportant or redundant features, which is especially useful when dealing with high-dimensional data. In the realm of data completion, sparse regression can help us predict possible values of missing data based on existing partial data, thereby completing the dataset, thus, sparse regression plays an extremely significant role in these domains (Erichson *et al.*, 2020). We incorporate this approach into the optimization of the forward modelling results in FWI:

$$\min_{\mathbf{U}} \frac{1}{2} \|\mathbf{AU} - \mathbf{D}\|_F^2 + \lambda \, \mathrm{R}(\mathbf{SU}), \tag{3}$$

where $\mathbf{A} \in \mathbb{R}^{m \times d}$ is the data-generating model for the real data $\mathbf{D}$, $\mathbf{U} \in \mathbb{R}^d$ is the recovery data, $\lambda$ is the regularisation parameter, $\mathrm{R}(\cdot)$ is any regularisation form and $\mathbf{S} \in \mathbb{R}^{n \times d}$ is a linear map.

In modern optimization, non-smooth regularisations are widely used in various applications. We innovatively apply it to seismic surveys to replace intensive acquisition with optimization while denoising the signal to achieve seismic data processing and waveform inversion at a lower cost (Gholami *et al.* 2021). The SR3 algorithm demonstrates more pronounced advantages compared to other regression algorithms (Zheng *et al.* 2019). One of its primary benefits is its capacity to effectively identify sparse signals, a significant effect brought about by the introduction of an auxiliary matrix. The application of such an auxiliary matrix becomes particularly crucial when the data contains noise and the model parameter matrix is ill-conditioned, and the presence of this matrix aims to aid us in more efficiently recovering the data matrix, reducing the artefacts induced by noise and data loss, thereby improving and optimizing the final inversion results (Aravkin *et al.* 2014).

Additionally, our approach offers flexibility, allowing us to adjust the regularisation function to accommodate various scenarios. Depending on the sparsity characteristics of different models, we can employ plural forms of regularisation functions. The flexibility in the selection of regularisation functions reflects the strong adaptability and flexibility of our approach, enabling us to achieve optimal results under a myriad of different settings and conditions of the constraint function and the fidelity of the relaxation problem (Esser *et al.* 2018). First, we need to construct an auxiliary variable $\mathbf{W}$ to relax equation 3:

$$\min_{\mathbf{U},\mathbf{W}} \frac{1}{2}\|\mathbf{AU}-\mathbf{D}\|_F^2 + \lambda\,\mathrm{R}(\mathbf{W}) + \frac{k}{2}\|\mathbf{W}-\mathbf{SU}\|_p, \tag{4}$$

where $\mathbf{W}^k = \mathrm{prox}_{\lambda k \mathrm{R}}(\mathbf{U}^k)$ is the auxiliary variable, which gradually approaches with $\mathbf{U}$; $\lambda$ is the penalty parameter, and $k$ is the relaxation parameter, where $k$ controls the degree of relaxation. The scalar $\|\cdot\|_p$ represents a different form of the regularisation used in the optimization, which can be flexibly used as different regularisation functions depending on the sparsity of the data matrix, such as $\ell_1$ and $\ell_2$ norm, even nuclear norm.

Associated with equation 4 is a form of the value function, which allows us to precisely depict this relaxed framework (Champion *et al.* 2020). A value function offers a quantitative representation of a problem, capable of revealing the nature and characteristics of the problem, aiding in our understanding of its essence, and providing direction for problem-solving. Specifically, this value function is obtained by minimizing equation 4 over $\mathbf{x}$, a process equivalent to finding a $\mathbf{x}$ that minimises equation 4. This minimum corresponds to the value function we are discussing, which to some extent reflects the properties of the optimal solution to the problem:

$$\mathbf{v}(\mathbf{W}) = \min_{\mathbf{U}} \frac{1}{2}\|\mathbf{AU}-\mathbf{D}\|_F^2 + \frac{k}{2}\|\mathbf{W}-\mathbf{SU}\|_p. \tag{5}$$

We assume that $\mathbf{H}_k = \mathbf{A}^\top\mathbf{A} + k\mathbf{S}^\top\mathbf{S}$ is invertible, so that $\mathbf{U}(\mathbf{W}) = \mathbf{H}_k^{-1}(\mathbf{A}^\top\mathbf{D} + k\mathbf{S}^\top\mathbf{W})$ is unique. And:

$$\mathbf{F}_k = \begin{bmatrix} k\mathbf{A}\mathbf{H}_k^{-1}\mathbf{S}^\top \\ \sqrt{k}(\mathbf{I}-k\mathbf{S}\mathbf{H}_k^{-1}\mathbf{S}^\top) \end{bmatrix}, \tag{6}$$

$$\mathbf{G}_k = \begin{bmatrix} \mathbf{I} - k\mathbf{A}\mathbf{H}_k^{-1}\mathbf{A}^\top \\ \sqrt{k}\mathbf{S}\mathbf{H}_k^{-1}\mathbf{A}^\top) \end{bmatrix}, \tag{7}$$

$$\mathbf{g}_k = \mathbf{G}_k \mathbf{D}, \tag{8}$$

which provides a closed-form expression for:

$$\mathbf{v}(\mathbf{W}) = \frac{1}{2}\|\mathbf{F}_k \mathbf{W} - \mathbf{g}_k\|_F^2. \tag{9}$$

Equation 4 then reduces to

$$\min_{\mathbf{W}} \frac{1}{2}\|\mathbf{F}_k \mathbf{W} - \mathbf{g}_k\|_F^2 + \lambda \mathrm{R}(\mathbf{W}). \tag{10}$$

To elucidate the superiority of the SR3 algorithm more accurately, we demonstrate partial minimization improves the condition of the problem in Figure 2. In Figure 2 (a), the coloured ellipses depict the contours of $\|\mathbf{AU}-\mathbf{D}\|_F^2$, while in Figure 2 (b), the contours of $\|\mathbf{F}_k \mathbf{W} - \mathbf{g}_k\|_F^2$ are vividly portrayed as a circle (Zheng et al. 2019). In section (a), we exhibit the contour lines of the quadratic part similar to the LASSO problem (coloured ellipses) and the approximate solution paths (red solid line) in horizontal projection. In section (b), the contour lines of the quadratic part of the SR3 loss function (coloured circle) and the corresponding approximate paths (red solid line) of the SR3 solution in the relaxed coordinates $\mathbf{W}$ are shown. Additionally, the grey diamonds indicate the contour lines of the $\ell_1$ norm of the LASSO problem in each coordinate group.

Figure 2 evidently shows that for the widely applied class of LASSO-like problems, the properties of $\mathbf{F}$ are generally superior to $\mathbf{A}$, particularly in terms of the condition number, where $\mathbf{F}$ is typically smaller than $\mathbf{A}$. The ratio of the maximum to minimum singular values of $\mathbf{F}$ is smaller, thereby compressing the contour lines into a shape closer to a sphere, which accelerates convergence and enhances performance. Moreover, executing proximal gradient descent solely in **W** naturally resolves these types of problems. The formulas for **F** and **G** can also be applied to acceleration methods, such as the FISTA algorithm. Overall, the SR3 algorithm reduces the singular values of F relative to **A** and has a weaker impact on small singular values. This effect "squeezes" the contour lines into a near-spherical shape, thereby accelerating convergence and enhancing performance (Zheng *et al*. 2018).

Finally, the algorithm requires two parameters to be specified simultaneously; the parameter $\lambda$ determines the strength of the regularisation, while *k* determines the degree of relaxation. We set the coefficient threshold to $\mathcal{T}$, then:

$$\lambda = \frac{\mathcal{T}^2 k}{2}. \tag{11}$$

With equation 11, we can change the two-parameter selection problem to a single-parameter selection and significantly improve the algorithm. In addition, we suggest a cross-validation approach to achieve parameter tuning of the automatic strategy.

## 3. Numerical examples

We utilised two types of homogeneous media and two distinct artificially synthesised velocity models to evaluate the performance of FWI optimised based on the Sparse Relaxation Regularised Regression (SR3) algorithm. The specific models include the single-layer homogeneous medium and double-layer homogeneous medium mentioned in this section, as well as the 2004 BP and SEG/EAGE Overthrust models. In our experiments, we employed the frequency-domain forward modelling algorithm and introduced the Perfectly Matched Layer (PML) as the absorption boundary condition (Pratt 1999).

Furthermore, FWI is notably susceptible to the interference of noise. To articulate the impact of noise on inversion outcomes quantitatively and to juxtapose the noise-reducing capabilities of distinct algorithms, we define the intensity of noise present in a signal utilizing the formula for Signal-to-Noise Ratio (SNR) (de Ridder & Dellinger 2011):

$$\text{SNR} = 20 * \text{Log}_{10}(\frac{\|\mathbf{D}\|_2}{\|\mathbf{E}\|_2}), \qquad (12)$$

where the $\mathbf{E}$ is the noise. To further illustrate the outlier suppression capabilities of the SR3-based FWI, we employ a signal with a low signal-to-noise ratio (SNR) of 10 *dB* across different models.

To evaluate the simulation accuracy of different algorithms for the model, we employ the model error:

$$\|\mathbf{M}_{true} - \mathbf{M}_{inv}\|_2 / \|\mathbf{M}_{true}\|_2, \qquad (13)$$

where the $\mathbf{M}_{true}$ and $\mathbf{M}_{inv}$ represent the true and inversion model velocity, respectively (Warner & Guasch 2016).

When implementing regression algorithms, we have several options to optimise our model. Specifically, we can consider using different regularises or a combination of various regularisation methods, which can help us achieve more effective optimization in equation 5. As demonstrated in Figure 1, we provide a three-dimensional visualization of two common regularises $\ell_1$ regularise in Figure 1 (a) and $\ell_2$ regularises in Figure 1 (b). Both regularises are widely applied in geophysics and engineering, and each possesses unique advantages. For the $\ell_1$ regularise, we can define it as the P regularisation in equation 5. The $\ell_1$ regularise emphasises the sparsity of regularisation, a feature that can greatly enhance the denoising capability of the algorithm, helping the model to focus on the most important features and, thus, improve the interpretability of the model. On the other hand, we could also select the $\ell_2$ norm as the P regularisation. The $\ell_2$ regularise emphasises the generalization capability of the algorithm. It can improve the smoothness and stability of the model and effectively prevent overfitting.

To demonstrate the core theory of the SR3 algorithm we adopted, we present in Figure 2 an illustrative diagram of the gradient iteration process using the LASSO problem as an example. Figure 2 (a) represents a schematic diagram of the update path of the proxy gradient under the action of the $\ell_1$ regularise for an elliptical contour loss function. Figure 2 (b) highlights the core idea of the SR3 algorithm, which "tightens" the elliptical contour of the loss function to an approximate circle, thereby accelerating the convergence speed and performance of regression computations. This accelerated convergence enables us to embed this algorithm into conventional FWI without the pressure of computational memory and time consumption.

**3.1 Single-layer homogeneous medium**

Firstly, the effectiveness of the SR3 algorithm in the application of high-frequency spatial mixing phenomena has been thoroughly tested in the context of a single-layer homogeneous medium. Figure 3 (a) shows the basic structure of a single-layer homogeneous medium. Figure 3 (b) provides a one-dimensional velocity model of a single-layer homogeneous medium that achieves a velocity of 2 km/s. Based on this, Figures 3 (c-e) further show the wavefield for the analytical solution of the monolayer homogeneous medium at frequencies of 10 Hz, 13 Hz, and 15 Hz. Figures 4 (a-d) show the results of the conventional algorithm for the monolayer homogeneous medium at 10 Hz. The discretized colour map is intended to improve recognition performance. Figures 4 (e-h) show the results after processing by the SR3 algorithm for the monolayer homogeneous medium at 10 Hz. To better compare the forward results at different frequencies in the case of the multi-scale algorithm and the processing effects of the SR3 algorithm, we have also compared the wavefield results and processing results at 13 Hz and 15 Hz, respectively. As shown in Figures 5 (a-h), the wavefield of the conventional algorithm at 13 Hz is compared with the wavefield processed by the SR3 algorithm in the same methods and order of comparisons as in Figures 4 (a-h). In addition, Figures 6 (a-h) show the comparison between the conventional wavefield results and the SR3 processed wavefield for the 15 Hz case, with the same methods and order of comparisons as in Figures 4 (a-h). The discretized colour map is intended to improve recognition performance.

**3.2 2004 BP Model**

Further, we test our proposed algorithm in synthetic data to verify its performance in near-real situations. Figure 7 (a) shows the real velocity model for the 2004 BP model, and Figure 7 (b) shows the initial velocity model used for the FWI with a linear velocity increase ranging from 1.5 km / s to 5 km / s. To test the effect of noise, we add random background noise to the recorded data set, as shown in Figure 8. The middle part of the 2004 BP model describes the geological characteristics of the Eastern / Central Gulf of Mexico and offshore Angola. Because the middle part of the model is composed of a high-velocity salt body, the division of the salt body is one of the difficulties in the inversion of the model. In addition, the channels are located near the salt body, which will further affect the velocity inversion and increase the difficulty of inversion (Billette & Brandsberg-Dahl 2005). The source-receiver wavefield of the 2004 BP model is demonstrated for the 3 Hz scenario, as illustrated in Figure 8 (a). Using this as a basis, we integrated 10 *dB* of random noise to mimic the interference that would typically arise in actual production. The noise equation is provided in equation 12. Figure 8 (b) depicts the 10 *dB* random noise, and Figure 8 (c) represents the wavefield after the clean wavefield has been linearly combined with the random noise. Consequently, Figure 8 (d) presents the subsampled wavefield, which we simulate as a low-quality wavefield influenced by downsampling. This low-quality wavefield is evenly distributed to incorporate the missing trace, which is then mixed with random noise and overlaid linearly onto the clean wavefield. The entire process results in the wavefield matrix containing 10 *dB* noise and the missing trace, as shown in Figure 8 (e). The subsequent FWI procedures hinge on the inversion iterations of the low-quality forward wavefield established through this process. We compute the source-receiver wavefield of the 2004 BP model within the frequency domain, as illustrated in Figure 9. In this figure, Figure 9 (a) presents the clean wavefield, while Figure 9 (b) offers a side view, adjusting the perspective from (a) to a view from (105,1). We follow the simulation steps mentioned above and linearly overlay the clean wavefield with random noise and down-sampling simulations, resulting in a low-quality wavefield depicted in Figure 9 (c). Figure 9 (d) provides a side view of this low-quality wavefield in (c). In the following step, we employ the SR3 optimization algorithm, as proposed in our method, to enhance the quality of the suboptimal wavefield, culminating in a superior version shown in Figure 9 (e). As can be seen, the continuity of the wavefield improves progressively from low to high frequency, effectively denoising and supplementing the matrix for reconstruction. Figure 9 (f) offers a side view of this improved wavefield in (e). In addition, we also give the side view of the simulated subsampled wavefield as in Figure 10, where (a) is the clean wavefield at 1 Hz, (b) is the low-mass wavefield obtained by simulated downsampling, and (c) is the wavefield optimised by SR3 for the low-mass wavefield, and our proposed algorithm significantly improves the continuity and uniformity of the wavefield. And (d-f) is the case at 2 Hz, and the conclusion is consistent with the above. Additionally, eigenvalues and eigenvectors are crucial for the dimensionality reduction of the wavefield matrix, as they can capture the most significant directions of variation in the matrix, making them extremely useful for interpolation and reconstruction algorithms. We present the results of singular value decomposition applied to the source-receiver wavefield at different frequencies in Figure 11. As can be seen, there are no extreme size differences in the eigenvalues of the wavefield matrix, but the eigenvalues decrease more rapidly at higher frequencies. Consequently, the reconstruction and interpolation of low-frequency matrices become particularly important. In other words, the optimization of low-frequency wavefield matrices can significantly enhance the accuracy and precision of the inversion. Following these, we present the inversion results of the multi-scale frequency domain FWI, as shown in Figure 12. The sub-figures 12 (a1-a8) represent the broadband inversion results based on the conventional Tikhonov regularisation FWI at frequencies of 1.20 Hz, 2.99 Hz, 3.58 Hz, 5.16 Hz, 7.43 Hz, 10.70 Hz, 12.84 Hz, and 15.41 Hz, respectively. The forward modelling data were linearly added with 10 *dB* of random background noise and missing trace, with the processing flow and the results shown in Figures 8 and 9. In addition, sub-figures 12 (b1-b8) represent the FWI inversion results obtained from the wavefield processed by SR3, with the same frequency range. To showcase the difference in performance, we calculated the differences between the velocity model obtained by the conventional algorithm and the true velocity model (c1-c8), as well as the differences between the FWI inversion results based on SR3 and the true velocity model (d1-d8), both using the same colour map. The 2004 BP model is a sharp contrast velocity model, where the velocity of the salt body part is significantly different from the surrounding velocity, making the delineation of the salt body contour a challenging aspect of inversion. Additionally, the structure of the water channels on both sides also poses a challenge for inversion. From Figure 12, we can see that the conventional FWI algorithm, affected by background noise and downsampling, has indistinct inversion results for the deep salt column structure and exhibits inversion anomalies for the structures of the water channels on both sides, especially an error in the structure on the left side. In contrast, the FWI optimised by the SR3 algorithm performs more excellently; the structure of the water channel on the left is correct and clear, the continuity of the two salt columns in the middle is better, and its velocity difference model compared with the true velocity is cleaner. Meanwhile, the velocity difference model of the conventional algorithm, compared with the true velocity, shows more distinct salt body contours and water channel structures, indicating it is not as clean in its processing. Figure 13 shows the convergence rates of the misfit error and model error for both methods, indicating that the convergence rate of SR3 (represented by the red solid line) is faster than that of the conventional FWI (represented by the blue solid line). Finally, we compare the one-dimensional velocity models of the two methods across six data sets, both horizontally and vertically. Figure 14 presents the comparison of velocities at six different $X$ locations. Anomalies and missing traces can lead to obvious errors in the inversion of the conventional FWI, such as a velocity anomaly at $X = 17.88$ km. However, the FWI based on the SR3 algorithm can effectively optimise and reconstruct the forward-modelled data, thereby correcting these velocity anomalies, which display a normal velocity curve, unimpacted by missing traces. Therefore, this demonstrates the superiority of our proposed algorithm over the conventional one. The comparison of lateral velocities is one of the more challenging aspects, as shown in Figure 15. Due to the propagation characteristics of seismic waves, the lateral velocity comparison of conventional FWI is often of inferior quality. However, the lateral velocity comparison of FWI, based on SR3 preprocessing, closely matches the true velocity model. This match reflects sufficient velocity compensation at the shallow part of the model and better control of anomalies at deeper parts. In contrast, the lateral velocity of the conventional algorithm, especially at the deeper part, still exhibits noticeable velocity anomalies due to the influence of noise and missing traces.

**3.3 SEG/EAGE Overthrust Model**

We proceeded to evaluate the performance of our refined approach using the SEG/EAGE Overthrust Model. This model is a widely adopted geological construct in seismic exploration, predominantly employed to characterise geological formations with overthrust structures. The P-wave velocity of the SEG/EAGE Overthrust Model, as depicted in Figure 16 (a), is widely utilised for testing and validating various geophysical algorithms (Yuan *et al.* 2015). The Overthrust Model exhibits numerous complex structures, inclusive of a plethora of thrust faults and fluvial sediments, which facilitates our exploration of the improved FWI's performance and representation under intricate geological conditions. Figure 16 (b) indicates the initial velocity model. Similar to the 2004 BP model, we present the clean wavefield of the SEG/EAGE Overthrust Model at 3 Hz, as depicted in Figure 17 (a), along with the 10 *dB* random noise shown in Figure 17 (b). The noise-included wavefield, Figure 17 (c), is obtained by linearly superimposing the random noise onto the clean wavefield. Subsequently, we simulate a subsampled wavefield by uniformly downsampling the matrix, resulting in Figure 17 (d). Finally, we superimpose this onto the clean wavefield to produce a subsampled wavefield matrix with 10 *dB* noise and missing trace, as represented in Figure 17 (e). The inversion for the SEG/EAGE Overthrust Model is based precisely on this, and the SR3 algorithm is also used to optimise the low-quality wavefield shown in Figure 17 (e). Similarly, we present a three-dimensional stack of the source-receiver wavefield, as depicted in Figure 18 (a), which represents the clean wavefield from 0.5 Hz to 5 Hz. Figure 18 (b) offers a clearer side view, obtained by adjusting the perspective from (155,20) to (105,5) in Figure 18 (a). Figure 18 (c) demonstrates the subsampled wavefield matrix stack with 10 *dB* noise and missing trace, obtained by following the two-step processing procedure from Figure 17. In response, we employ the proposed SR3 to optimise it, resulting in a superior wavefield stack, as shown in Figure 18 (e). Our proposed algorithm yields commendable matrix interpolation and reconstruction results in both the 2004 BP model and the Overthrust model, exhibiting improved continuity in the wavefield stack, superior noise suppression, and clearer wavefield details. Then, we present the inversion results of the multi-scale FWI. Figure 19 shows images (a1-a8) depict the inversion results obtained using conventional Tikhonov regularisation FWI, whereas Figures (b1-b8) exhibit the inversion results procured via FWI based on SR3 optimization. The results correspond to the inversions at frequencies of 2.15 Hz, 3.09 Hz, 5.35 Hz, 7.70 Hz, 9.24 Hz, 11.09 Hz, 13.31 Hz, and 15.97 Hz, respectively. From the results, it is evident that the enhanced algorithm distinctly outperforms in handling complex structures. It not only delineates shallow layered structures more clearly, enhancing their contours, but also sketches and outlines velocity disparities more effectively at depth, avoiding the blurry deep structures often obtained from conventional algorithms. To emphasise the differences, we calculated the discrepancies between the two algorithms and the true velocity model, as shown in (c1-c8) and (d1-d8) of Figure 19. Under the same colourmap conditions, the discrepancy between the inversion results of the improved algorithm and the true velocity model is visibly superior, exhibiting indistinct layering of the discrepancy, uniform velocity differences between layers, and no notable velocity anomalies at the bottom. Moreover, the improved algorithm ensures the stability of FWI in the high-frequency range without inversion anomalies caused by the spatial aliasing effect and always maintains stability and accuracy. These indeed validate the advantages and efficacy of the improved algorithm. We give the comparison of the misfit error and model error of the two algorithms in Figure 20, where the model error is given by equation 13. Lastly, we present the comparison of the one-dimensional velocity models derived from the two algorithms at various $X$ and $Y$ positions, representing the vertical and horizontal dimensions, respectively. In Figure 21, velocity model discrepancies at six different $X$-positions are presented, specifically at 3.13 km, 4.17 km, 7.77 km, 10.97 km, 17.67 km, and 18.33 km. From the curves, it is apparent that the velocity model generated by the improved algorithm conforms more closely to the true velocity at many locations with substantial velocity disparities. In contrast, the conventional algorithm often performs poorly in in-depth fitting, sometimes resulting in inversion errors. Figure 22 presents the more challenging cross-sectional velocity comparisons. Due to the propagation nature of seismic waves and the constraints of data acquisition, seismic waves emanate spherically from the source, *i.e.*, they cover a greater vertical distance compared to the horizontal distance. Moreover, in practical seismic exploration, due to economic and implementation restrictions, we typically cannot place a receiver at every point on the ground. Results in a non-uniform data sampling on the surface, affecting the accuracy of lateral velocity determination. Despite these challenges, the velocity model obtained from the improved algorithm in Figure 22 still fits the real velocity model better. We present data for six groups, respectively: (a) $Y = 0.40$ km; (b) $Y = 0.81$ km; (c) $Y = 1.66$ km; (d) $Y = 3.07$ km; (e) $Y = 3.44$ km; (f) $Y = 3.68$ km. In every group, the FWI based on SR3 optimization demonstrates superior performance. The conventional method is affected by the inferior quality of the wavefield data, which includes noise and missing traces, and fails to meet the challenges under extreme conditions. In contrast, our proposed new algorithm successfully addresses these challenges, thus demonstrating its effectiveness, robustness, and practicality.

## 4. Discussions

FWI, as a research direction that has received much attention, many scholars have made outstanding contributions to this field. However, the truth is that, even though it has been proposed for several decades, it has not been widely applied due to certain shortcomings and defects. As an algorithm composed of multiple steps, in FWI, good forward modelling does not necessarily depend on good inversion, but good inversion certainly depends on good forward modelling. Therefore, the results of forward modelling are crucial for FWI. In practical production, it is well known that external factors can create many troubles and difficulties for data collection and data quality. Interference from noise, limitations on the azimuth angle, and other factors can all gradually reduce the quality of the data, thereby reducing the accuracy of forward modelling and increasing its difficulty. On the other hand, due to the large scale of seismic exploration, FWI, which relies on grid methods, has high demands on computer memory. Under the current situation where the theory of grid methods and computer memory are close to their limit, how to improve the quality of the forward-modelled wavefield and lay a better foundation for inversion is an urgent problem to be solved in FWI.

This paper proposes a preprocessing step before optimization, the Sparse Relaxed Regularised Regression (SR3) algorithm. This algorithm can adequately complete and denoise seismic wavefields without affecting the overall computational efficiency and memory consumption through the empowerment of an auxiliary matrix and the idea of compressing singular value space to speed up the convergence of the regression algorithm. We conducted tests in homogeneous media and two types of artificially synthesised data, and the results obtained were in line with our expectations.

The reason for proposing such an approach is precisely to solve the numerous problems encountered in the practical application of FWI. For example, FWI is often affected by data noise, so how to effectively denoise becomes a problem. The proposed algorithm can effectively denoise from low to high frequencies, allowing FWI to achieve denoising effects without relying on regularisation algorithms. Furthermore, the concept of full azimuth is often mentioned in FWI. Although this is an excellent system to improve FWI resolution, we have to consider the issues of cost-effectiveness and practicality. Full azimuth necessarily brings higher time and financial costs, contrary to the original intention of proposing FWI. The original reason for proposing FWI was to reduce time and financial costs through more innovative algorithms. Therefore, updating algorithms within FWI may be a more cost-effective way. Thus, the method of completing or reconstructing low-quality data through regression algorithm-like methods has an excellent research background and significance. The SR3 algorithm proposed in this paper is a higher-level sparse regression algorithm, mainly reflected in three aspects. (1) For the denoising problem of FWI, the SR3 algorithm has already been completed in preprocessing, so it is not necessary to carry it out during the minimization process, which avoids the computational pressure brought about by the need to introduce the model increment twice during the minimization process. (2) The SR3 algorithm is also essentially a multi-constraint optimization problem, so a composite regularise can be used, making the algorithm very flexible and robust. (3) Compared with other regression algorithms, the SR3 algorithm tightens the singular value space, as shown in Figure 2, greatly speeding up computational efficiency, without bringing computational pressure to the overall FWI.

However, although the SR3 algorithm is more advanced, as with other nonlinear algorithms, the selection of parameters is a difficult point. We do not object to the use of heuristic parameter selection methods, but a more logical and reasonable hyperparameter selection method should be considered. Although cross-validation, the L-curve method, etc., are all excellent solutions, how to choose the appropriate hyperparameters without increasing computational time and without the need for repeated computations is one of the contents worthies of deeper future research.

## 5. Conclusions

Our research introduces a more comprehensive approach to broadband FWI, integrating a preprocessing procedure that focuses on handling frequency domain source-receiver wavefields. This process is achieved through the application of the Sparse Relaxed Regularised Regression (SR3) algorithm. We propose a method designed to confront several of the most familiar challenges in seismic data processing: denoising, data reconstruction, and resistance to spatial aliasing effects. In the numerical part, we initiate our analysis by comparing the treatment effects of two homogeneous media cases: single-layer and double-layer media, respectively. We specifically highlight the variance between the wavefield before and after processing and the numerical solution in the high-frequency part. Subsequently, our proposed algorithm was applied to two sets of credible benchmarks widely used for further testing. The results from the forward modelling and inversion processes compellingly demonstrate the efficacy of our algorithm, with prominent features including significant optimization of noisy data and more comprehensive inversion of model details. In practical operations, the quality of seismic data encountered often parallels the quality of the simulation forward modelling data used in our research, which is typically low. In these extremely challenging situations, handling such low-quality seismic data becomes exceedingly difficult. Thus, the new optimization algorithm proposed in our research represents a promising and more rational approach for the further industrial application of FWI. It offers a potential pathway to significantly improve the outcomes of seismic inversion, especially in settings where the quality of the raw data may be suboptimal.

**Acknowledgements**

*Conflict of interest statement:* Authors declare that there is no conflict of interest.

## List of Figures

**Figure 1.** Three-dimensional visualization of two common regularises, (a) $\ell_1$ regularise and (b) $\ell_2$ regularise.

**Figure 2.** Illustrative figures of the gradient iteration process using the LASSO problem as an example, (a) conventional proxy-gradient process, (b) SR3 "tightens" the elliptical contour of the loss function to an approximate circle, thereby accelerating the convergence speed and performance of regression computations. The grey diamond is the contour of the $\ell_1$ norm, The solid red line is the direction of the update of the iteration.

**Figure 3.** (a) The basic structure of a single-layer homogeneous medium, (b) velocity model of the single-layer homogeneous medium which achieves a velocity of 2 km/s, (c-e) the wavefield model for the analytical solution of the monolayer homogeneous medium at frequencies of 10 Hz, 13 Hz, and 15 Hz. The horizontal axis is the offset distance, and the vertical axis is the depth.

**Figure 4.** (a-d) the results and comparisons of the conventional algorithm at 10 Hz, where (a) shows the numerical solution for the monolayer homogeneous medium at 10 Hz, (b) the real part of the difference between the numerical solution (Fig. 4 (a)) and the analytical solution (Fig. 3 (c)), (c) the real part of the ratio of the analytical solution to the numerical solution, (d) the angle part of the ratio of the difference (Fig. 4 (b)) to the analytic solution. (e-h) show the results after processing by the SR3 algorithm, where (e) shows the numerical solution after SR3 processed for the monolayer homogeneous medium at 10 Hz, (f) the real part of the difference between the numerical solution (Fig. 4 (e)) and the analytical solution (Fig. 3 (c)), (g) the real part of the ratio of the analytical solution to the numerical solution, (h) the angle part of the ratio of the difference (Fig. 4 (f)) to the analytic solution. The discretized colour map is intended to improve

recognition performance.

**Figure 5.** (a-d) the results and comparisons of the conventional algorithm at 13 Hz, with the same methods and order of comparisons as in Figure 4(a-d). (e-h) the results after processing by the SR3 algorithm at 13 Hz, with the same ordering as in Figure 4(e-h). The discretized colour map is intended to improve recognition performance.

**Figure 6.** (a-d) The results and comparisons of the conventional algorithm at 15 Hz, with the same methods and order of comparisons as in Figure 4(a-d). (e-h) the results after processing by the SR3 algorithm at 15 Hz, with the same ordering as in Figure 4(e-h). The discretized colour map is intended to improve recognition performance.

**Figure 7.** 2004 BP model, (a) true velocity model, (b) initial velocity model.

**Figure 8.** Source-receiver domain data set at 3 Hz of the 2004 BP model, the real part of the (a) clean data matrix, (b) 10 *dB* random noise, (c) wavefield matrix with 10 *dB* noise, (d) missing-trace matrix, (e) subsampled wavefield matrix with 10 *dB* noise and missing trace.

**Figure 9.** Three-dimensional low-frequency source-receiver wavefield from 0.5 Hz to 5 Hz of the 2004 BP model, (a) clean data matrix, (b) side view of (a); (c) subsampled wavefield matrix with missing-trace and 10 *dB* noise, (d) side view of (c); (e) wavefield matrix after SR3 optimization for (c), (f) side view of (e).

**Figure 10.** Three-dimensional side view low-frequency source-receiver wavefield; 1 Hz (a) clean wavefield, (b) subsampled wavefield with missing trace, (c) SR3 processed reconstructing wavefield; 2 Hz (d) clean wavefield, (e) subsampled wavefield with missing-trace, (f) SR3 processed reconstructing wavefield.

**Figure 11.** Eigenvalues of wavefield matrices with different frequencies after singular value decomposition processing. The horizontal axis is the eigenvalue index, the vertical axis is the normalised eigenvalue.

**Figure 12.** 2004 BP model inversion results, (a1-a8) Tikhonov regularisation FWI inversion results in 1.20 Hz, 2.99 Hz, 3.58 Hz, 5.16 Hz, 7.43 Hz, 10.70 Hz, 12.84 Hz, and 15.41 Hz, respectively, (b1-b8) FWI based on SR3 algorithm optimization inversion results in 1.20 Hz, 2.99 Hz, 3.58 Hz, 5.16 Hz, 7.43 Hz, 10.70 Hz, 12.84 Hz, and 15.41 Hz, respectively, (c1-c8) differences between the Tikhonov FWI and the true velocity model, (d1-d8) differences between the SR3-based FWI and the true velocity model.

**Figure 13.** Comparison of SR3 algorithm-based FWI with conventional Tikhonov regularisation-based FWI for quantification, (a) misfit error, (b) model error. The horizontal axis is the number of iterations, and the vertical axis is the error value.

**Figure 14.** 2004 BP model, one-dimensional velocity models at different X-positions, (a) X = 0.56 km; (b) X = 7.60 km; (c) X = 8.08 km; (d) X = 9.00 km; (e) X = 11.08 km; (f) X = 17.88 km, the vertical comparison of the actual velocity model (solid black line), initial velocity model (grey dotted line), the Tikhonov regularisation FWI velocity model (solid blue line), and the SR3-based FWI velocity model (solid red line).

**Figure 15.** 2004 BP model, one-dimensional velocity models at different *Y*-positions, (a) *Y* = 1.49 km; (b) *Y* = 2.01 km; (c) *Y* = 4.21 km; (d) *Y* = 4.43 km; (e) *Y* = 4.62 km; (f) *Y* = 4.99 km, the horizontal comparison of the actual velocity model (solid black line), initial velocity model (grey dotted line), the Tikhonov regularisation FWI velocity model (solid blue line), and the SR3-based FWI velocity model (solid red line).

**Figure 16.** SEG/EAGE overthrust model, (a) true velocity model. (b) initial velocity model.

**Figure 17.** Source-receiver domain data set at 3 Hz of the SEG/EAGE overthrust model, the real part of the (a) clean data matrix, (b) 10 *dB* random noise, (c) wavefield matrix with 10 *dB* noise, (d) missing-trace matrix, (e) subsampled wavefield matrix with 10 *dB* noise and missing trace.

**Figure 18.** Three-dimensional low-frequency source-receiver wavefield from 0.5 Hz to 5 Hz of the SEG/EAGE overthrust model, (a) clean data matrix, (b) side view of (a); (c) subsampled wavefield matrix with missing-trace and 10 *dB* noise, (d) side view of (c); (e) wavefield matrix after SR3 optimization for (c), (f) side view of (e).

**Figure 19.** The SEG/EAGE overthrust model inversion results. (a1-a8) Tikhonov regularisation FWI inversion results in 2.15 Hz, 3.09 Hz, 5.35 Hz, 7.70 Hz, 9.24 Hz, 11.09 Hz, 13.31 Hz, and 15.97 Hz, respectively; (b1-b8) FWI based on SR3 algorithm inversion results in 2.15 Hz, 3.09 Hz, 5.35 Hz, 7.70 Hz, 9.24 Hz, 11.09 Hz, 13.31 Hz, and 15.97 Hz, respectively; (c1-c8) differences between the Tikhonov FWI and the true velocity model; (d1-d8) differences between the SR3 FWI and the true velocity model.

**Figure 20.** Comparison of SR3 algorithm-based FWI with conventional Tikhonov regularisation-based FWI for quantification, (a) normalised misfit error, (b) model error. The horizontal axis is the number of iterations, and the vertical axis is the error value.

**Figure 21.** The SEG/EAGE overthrust model, 1-D velocity models at different *X*-positions, (a) *X* = 3.13 km; (b) *X* = 4.17 km; (c) *X* = 7.77 km; (d) *X* = 10.97 km; (e) *X* = 17.67 km; (f) *X* = 18.33 km. The vertical comparison of the actual velocity model (solid black line), initial velocity model (grey dotted line), the Tikhonov regularisation FWI velocity model (solid blue line), and the SR3-based FWI velocity model (solid red line).

**Figure 22.** The SEG/EAGE overthrust model, 1-D velocity models at different *Y*-positions, (a) *Y* = 0.40 km; (b) *Y* = 0.81 km; (c) *Y* = 1.66 km; (d) *Y* = 3.07 km; (e) *Y* = 3.44 km; (f) *Y* = 3.68 km. The horizontal comparison of the actual velocity model (solid black line), initial velocity model (grey dotted line), the Tikhonov regularisation FWI velocity model (solid blue line), and the SR3-based FWI velocity model (solid red line).

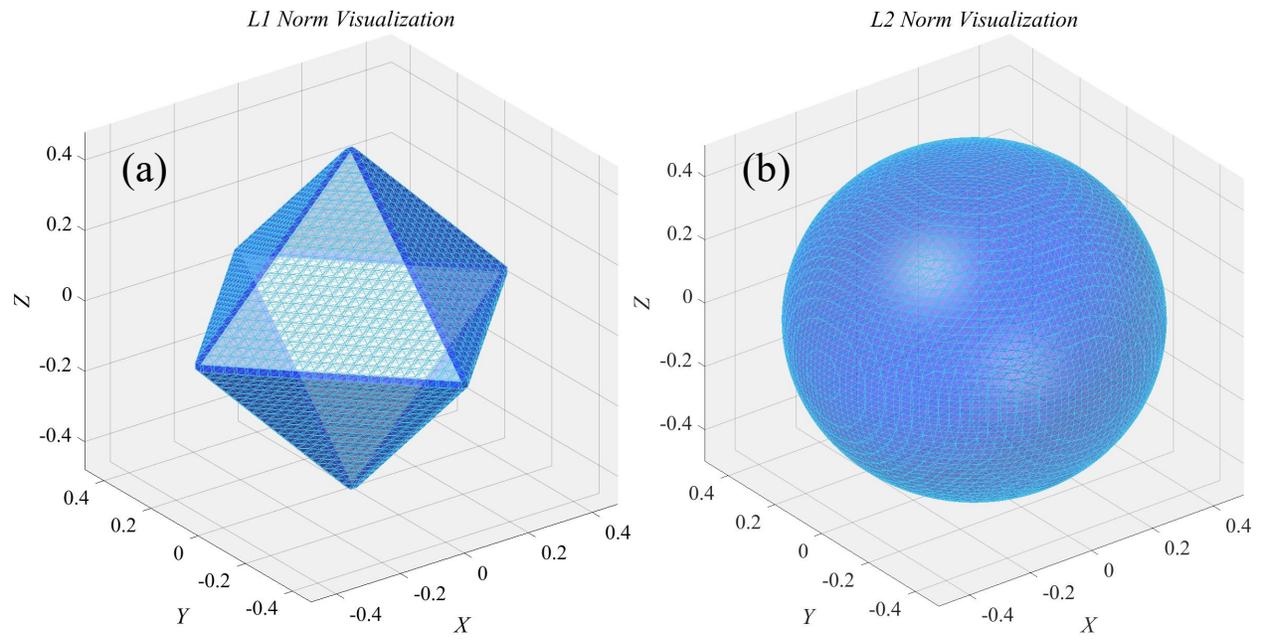

**Figure 1.** Three-dimensional visualization of two common regularises, (a) $\ell_1$ regularise and (b) $\ell_2$ regularise.

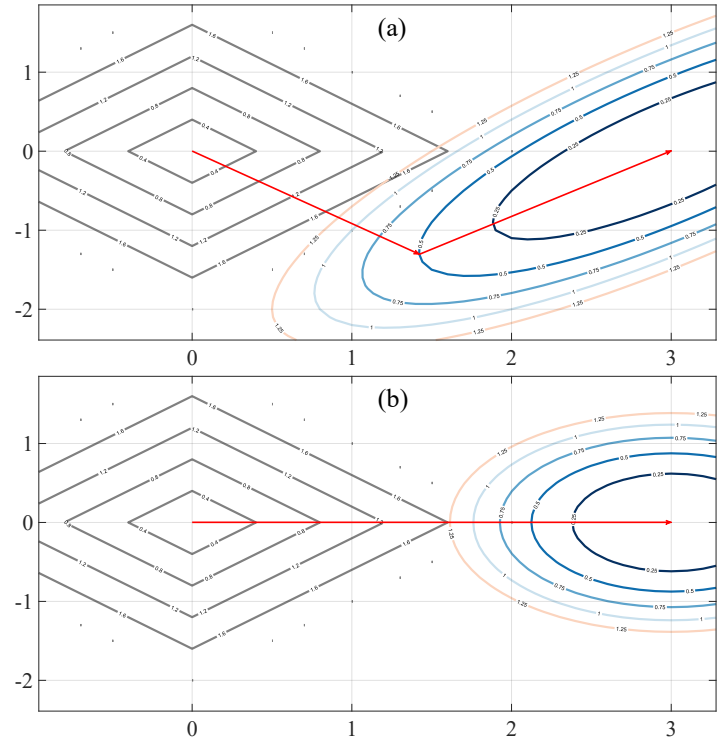

**Figure 2.** Illustrative figures of the gradient iteration process using the LASSO problem as an example, (a) conventional proxy-gradient process, (b) SR3 "tightens" the elliptical contour of the loss function to an approximate circle, thereby accelerating the convergence speed and performance of regression computations. The grey diamond is the contour of the $\ell_1$ norm, The solid red line is the direction of the update of the iteration.

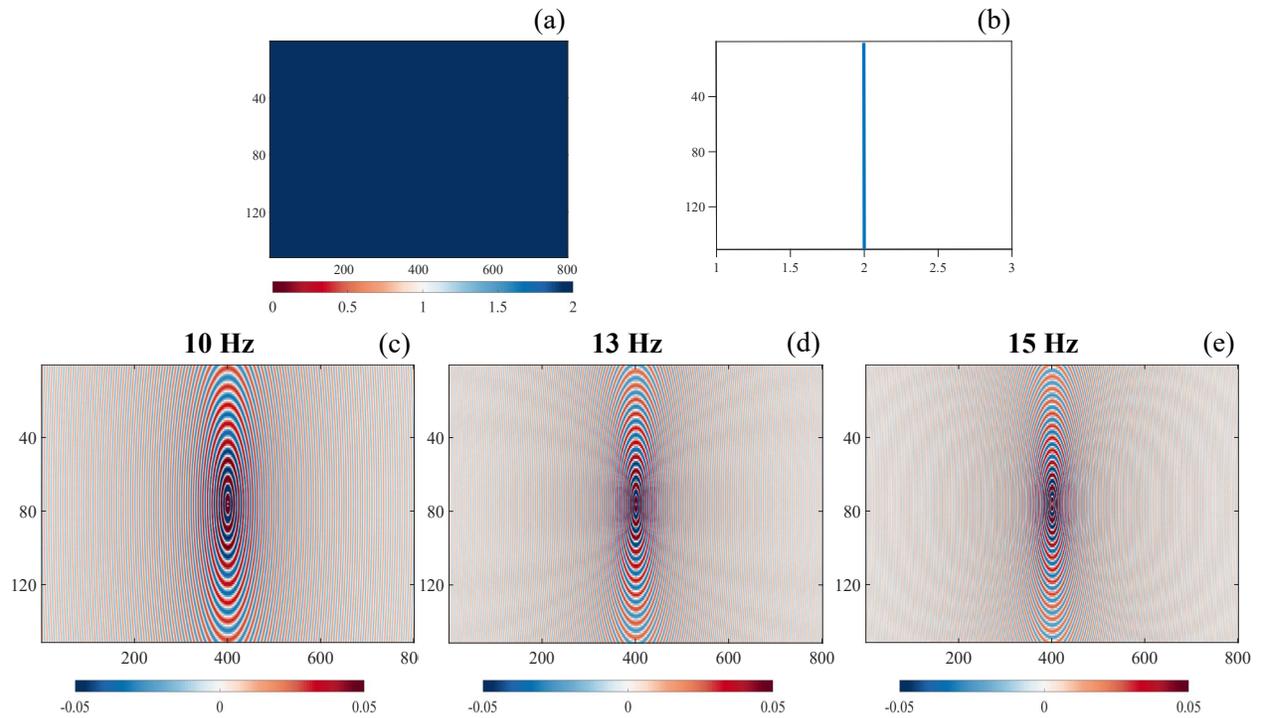

**Figure 3.** (a) The basic structure of a single-layer homogeneous medium, (b) velocity model of the single-layer homogeneous medium which achieves a velocity of 2 km/s, (c-e) the wavefield model for the analytical solution of the monolayer homogeneous medium at frequencies of 10 Hz, 13 Hz, and 15 Hz. The horizontal axis is the offset distance, and the vertical axis is the depth.

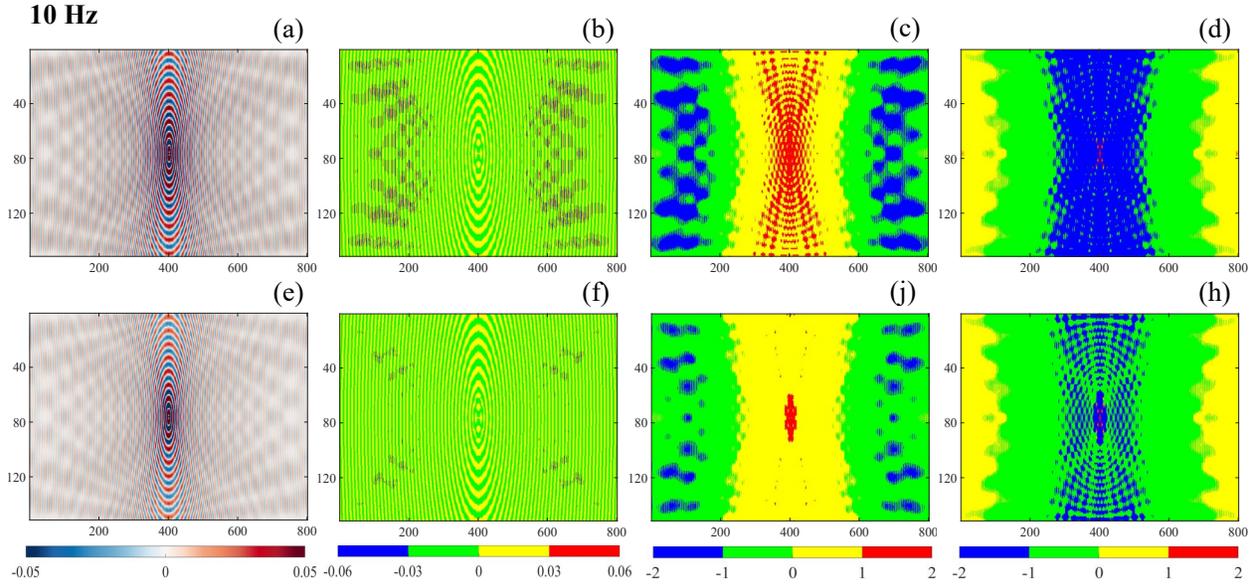

**Figure 4.** (a-d) the results and comparisons of the conventional algorithm at 10 Hz, where (a) shows the numerical solution for the monolayer homogeneous medium at 10 Hz, (b) the real part of the difference between the numerical solution (Fig. 4(a)) and the analytical solution (Fig. 3(c)), (c) the real part of the ratio of the analytical solution to the numerical solution, (d) the angle part of the ratio of the difference (Fig. 4(b)) to the analytic solution. (e-h) show the results after processing by the SR3 algorithm, where (e) shows the numerical solution after SR3 processed for the monolayer homogeneous medium at 10 Hz, (f) the real part of the difference between the numerical solution (Fig. 4(e)) and the analytical solution (Fig. 3(c)), (g) the real part of the ratio of the analytical solution to the numerical solution, (h) the angle part of the ratio of the difference (Fig. 4(f)) to the analytic solution. The discretized colour map is intended to improve recognition performance.

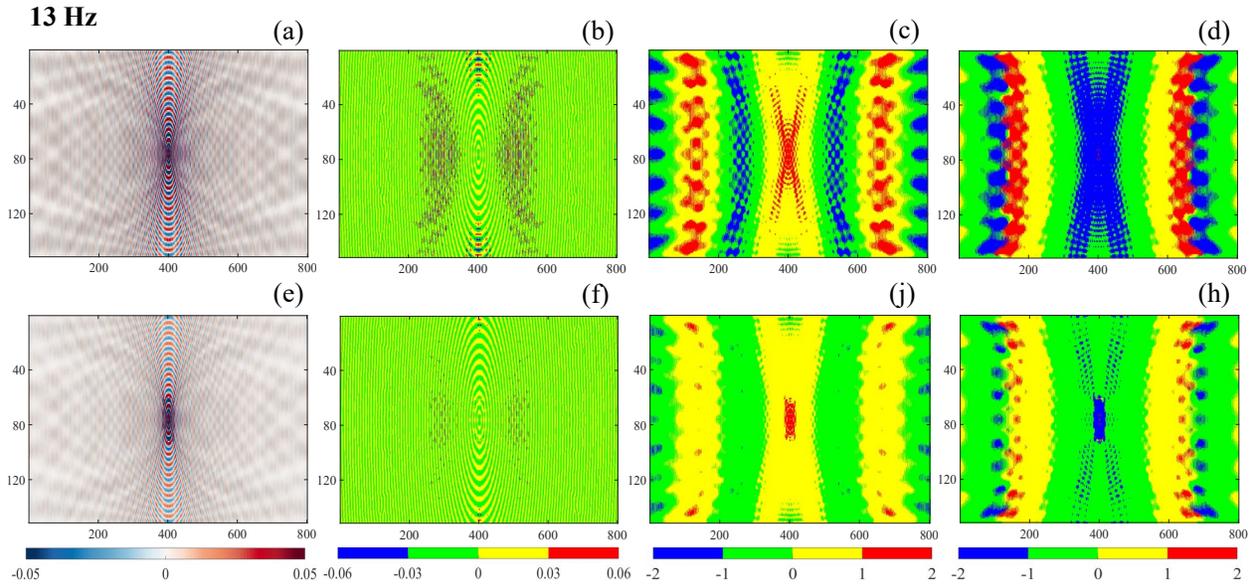

**Figure 5.** (a-d) the results and comparisons of the conventional algorithm at 13 Hz, with the same methods and order of comparisons as in Figure 4(a-d). (e-h) the results after processing by the SR3 algorithm at 13 Hz, with the same ordering as in Figure 4(e-h). The discretized colour map is intended to improve recognition performance.

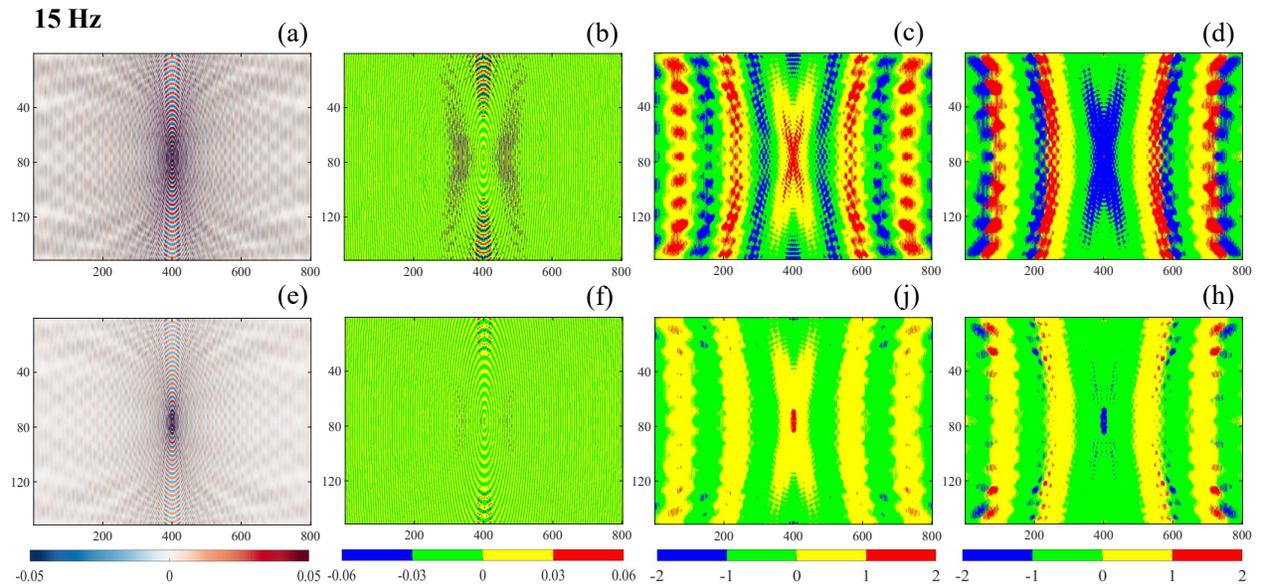

**Figure 6.** (a-d) The results and comparisons of the conventional algorithm at 15 Hz, with the same methods and order of comparisons as in Figure 4(a-d). (e-h) the results after processing by the SR3 algorithm at 15 Hz, with the same ordering as in Figure 4(e-h). The discretized colour map is intended to improve recognition performance.

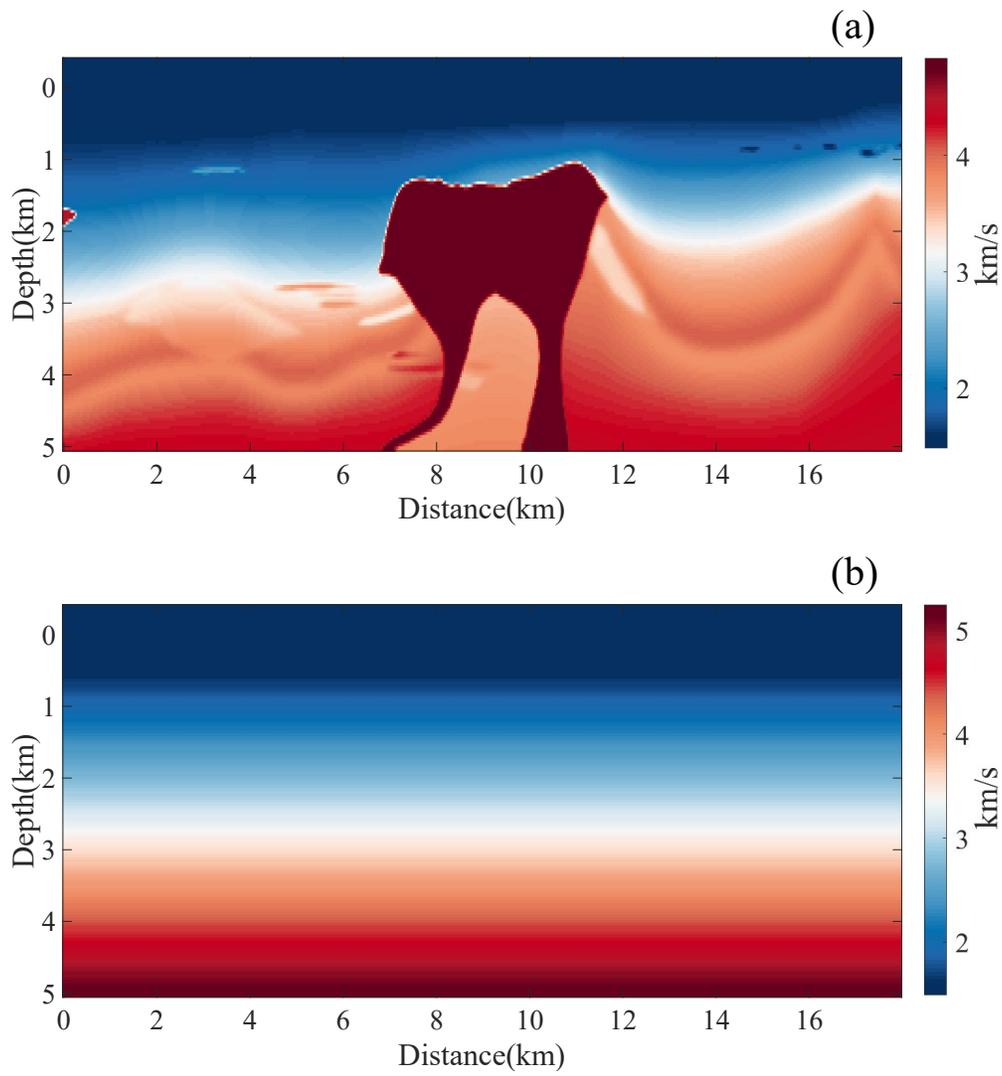

**Figure 7.** 2004 BP model, (a) true velocity model, (b) initial velocity model.

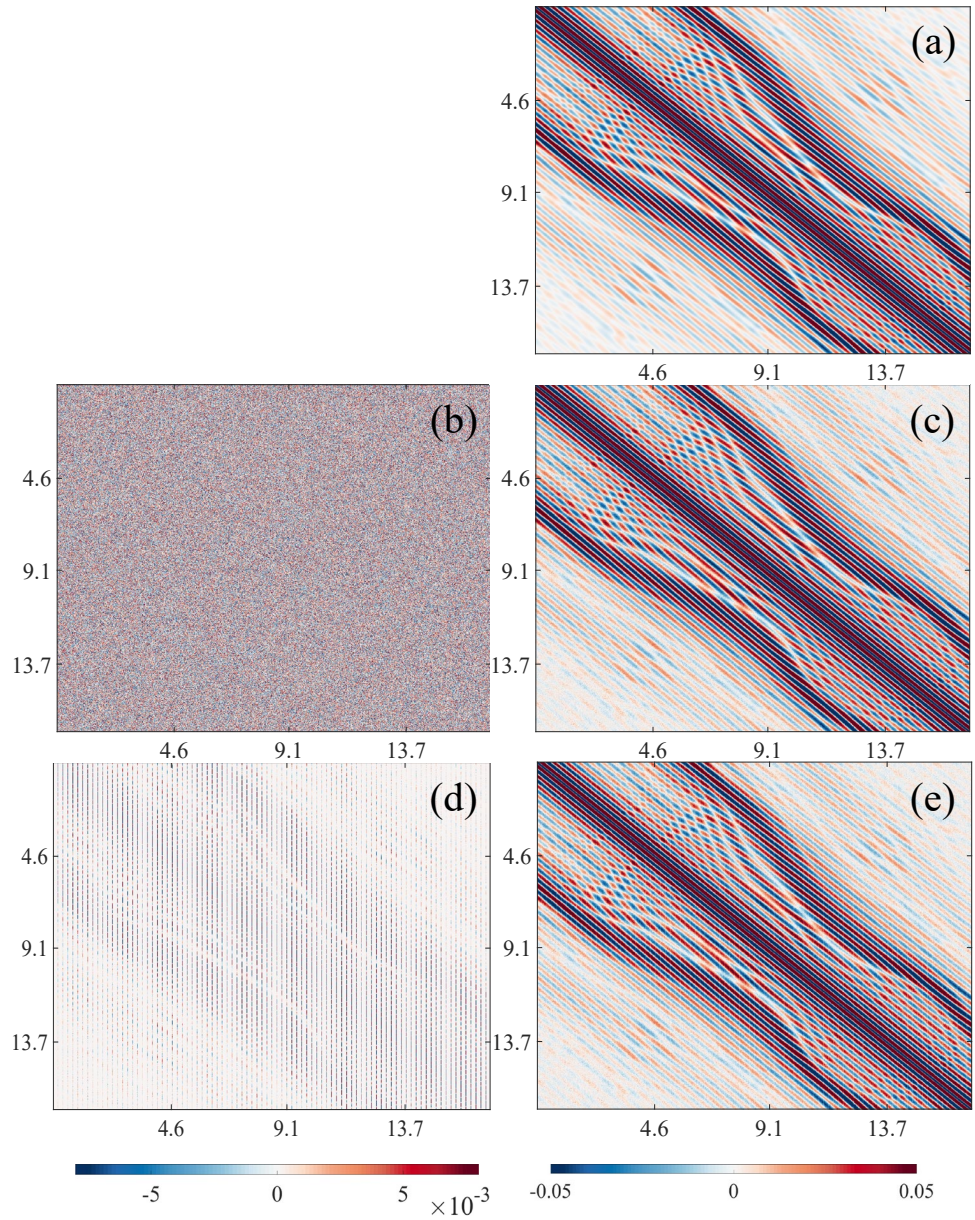

**Figure 8.** Source-receiver domain data set at 3 Hz of the 2004 BP model, the real part of the (a) clean data matrix, (b) 10 *dB* random noise, (c) wavefield matrix with 10 *dB* noise, (d) missing-trace matrix, (e) subsampled wavefield matrix with 10 *dB* noise and missing trace.

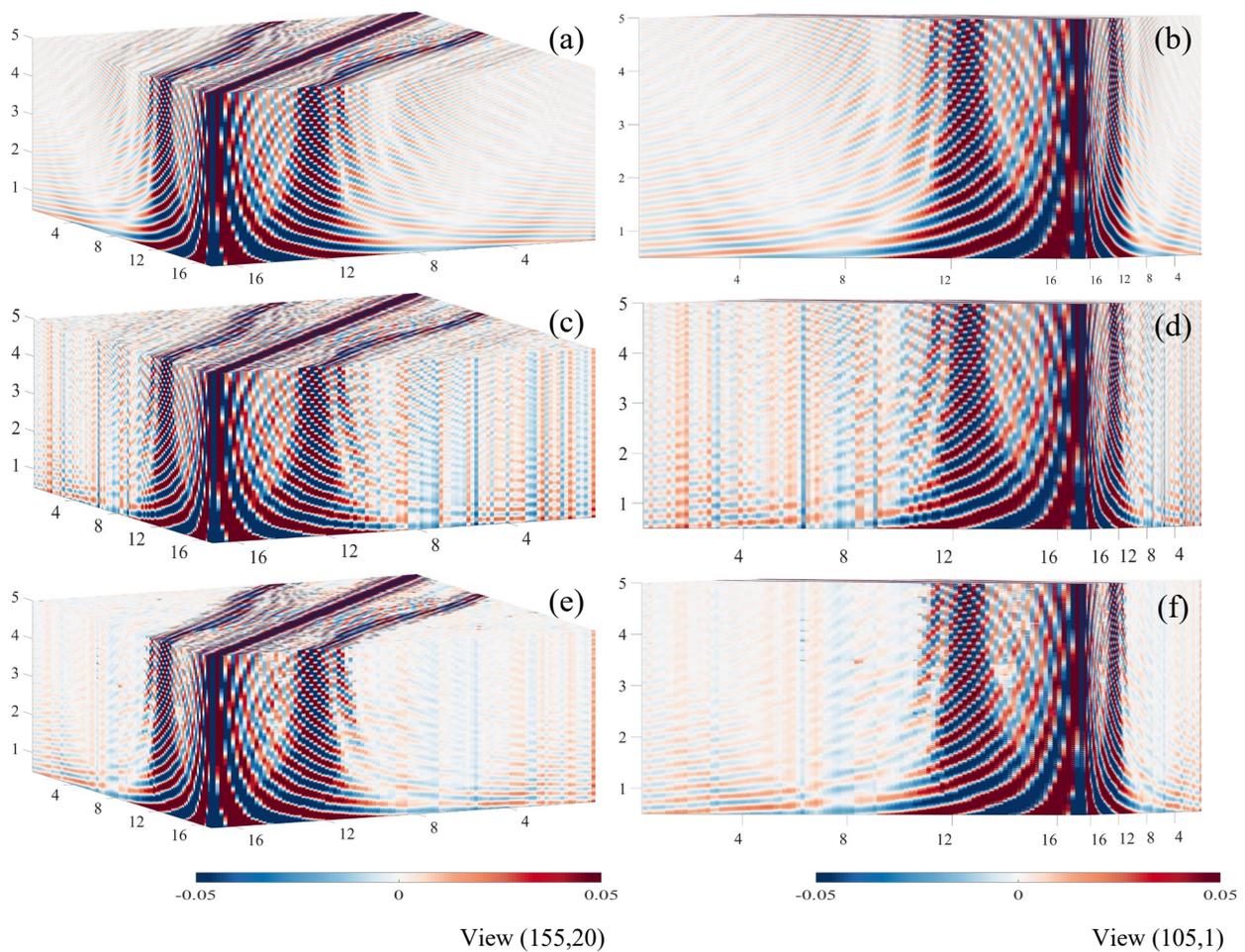

**Figure 9.** Three-dimensional low-frequency source-receiver wavefield from 0.5 Hz to 5 Hz of the 2004 BP model, (a) clean data matrix, (b) side view of (a); (c) subsampled wavefield matrix with missing-trace and 10 *dB* noise, (d) side view of (c); (e) wavefield matrix after SR3 optimization for (c), (f) side view of (e).

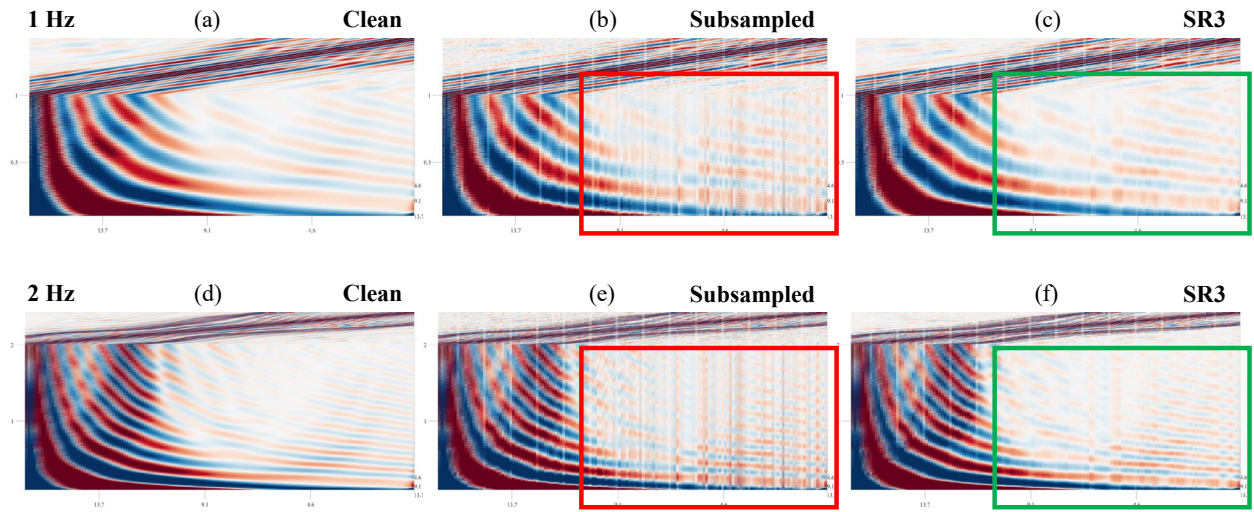

**Figure 10.** Three-dimensional side view low-frequency source-receiver wavefield; 1 Hz (a) clean wavefield, (b) subsampled wavefield with missing trace, (c) SR3 processed reconstructing wavefield; 2 Hz (d) clean wavefield, (e) subsampled wavefield with missing-trace, (f) SR3 processed reconstructing wavefield.

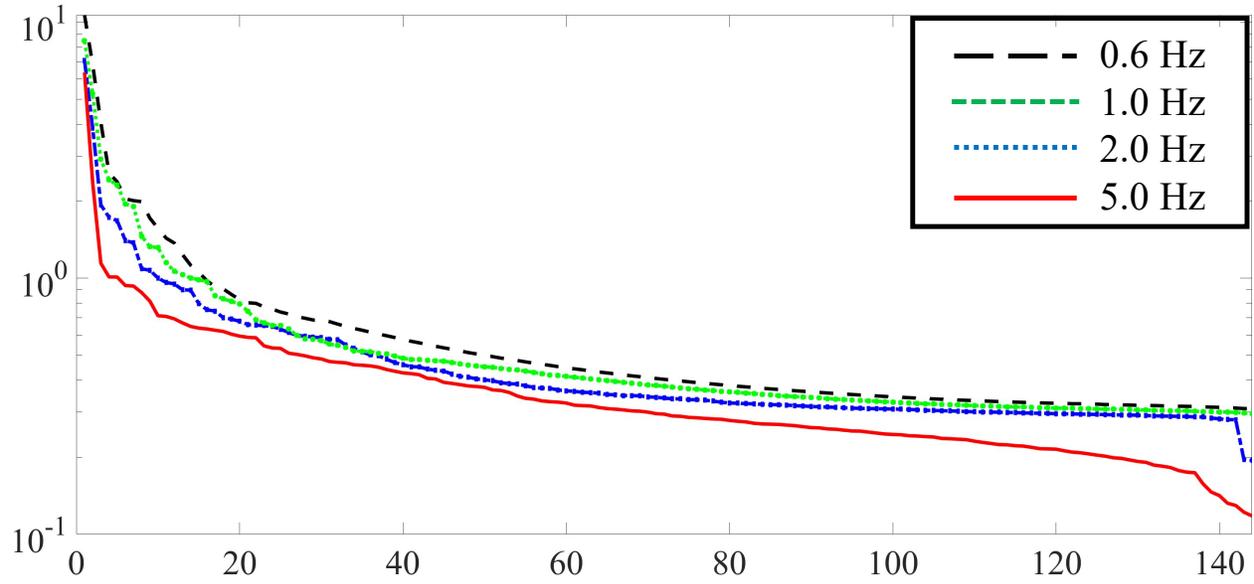

**Figure 11.** Eigenvalues of wavefield matrices with different frequencies after singular value decomposition processing. The horizontal axis is the eigenvalue index, the vertical axis is the normalised eigenvalue.

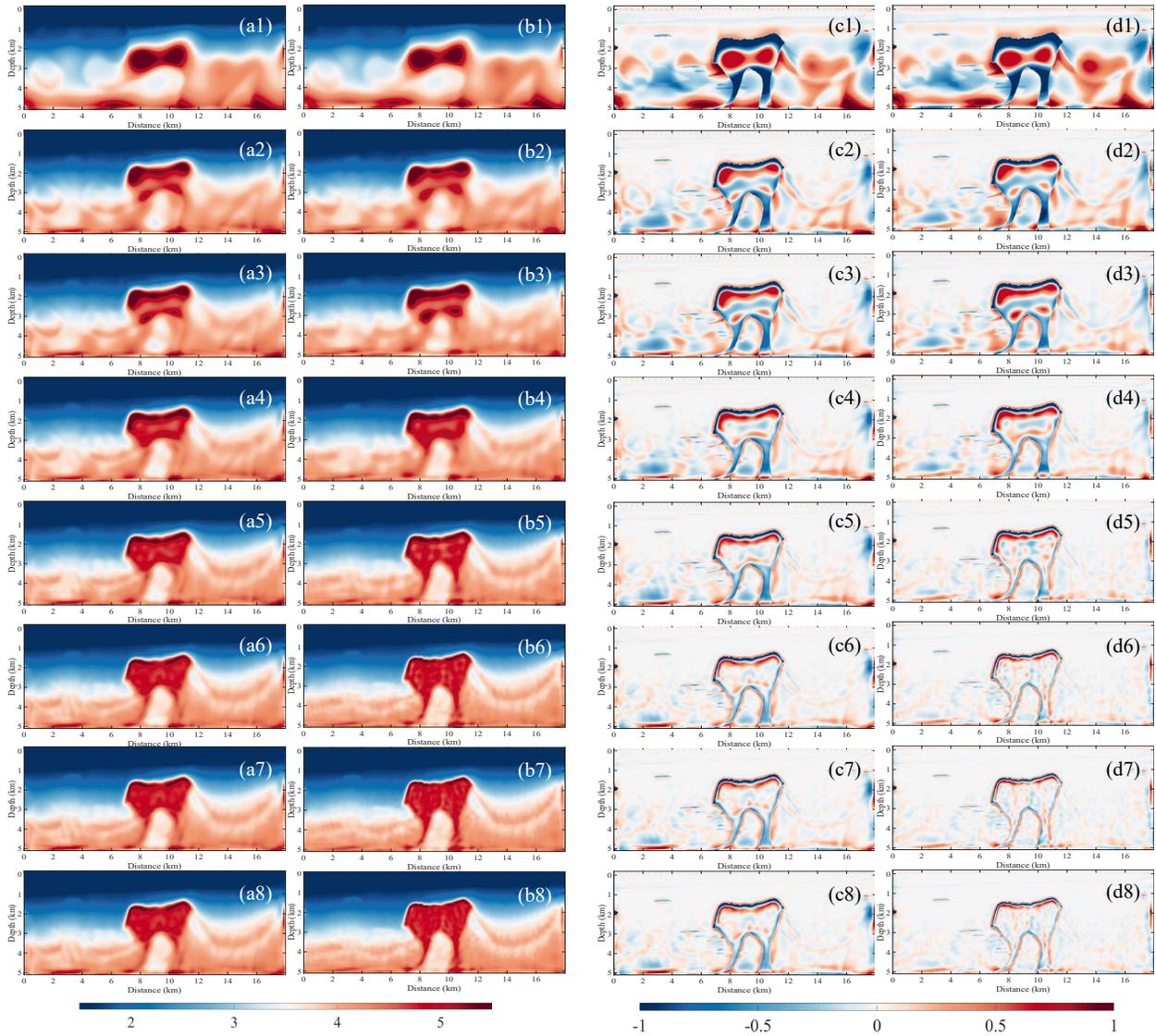

**Figure 12.** 2004 BP model inversion results, (a1-a8) Tikhonov regularisation FWI inversion results in 1.20 Hz, 2.99 Hz, 3.58 Hz, 5.16 Hz, 7.43 Hz, 10.70 Hz, 12.84 Hz, and 15.41 Hz, respectively, (b1-b8) FWI based on SR3 algorithm optimization inversion results in 1.20 Hz, 2.99 Hz, 3.58 Hz, 5.16 Hz, 7.43 Hz, 10.70 Hz, 12.84 Hz, and 15.41 Hz, respectively, (c1-c8) differences between the Tikhonov FWI and the true velocity model, (d1-d8) differences between the SR3-based FWI and the true velocity model.

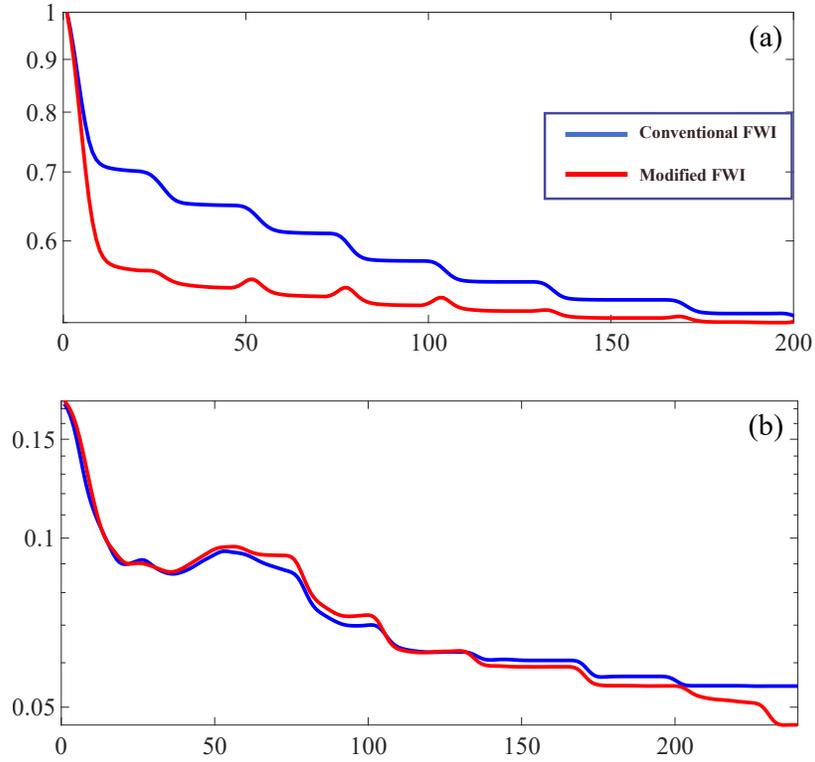

**Figure 13.** Comparison of SR3 algorithm-based FWI with conventional Tikhonov regularisation-based FWI for quantification, (a) misfit error, (b) model error. The horizontal axis is the number of iterations, and the vertical axis is the error value.

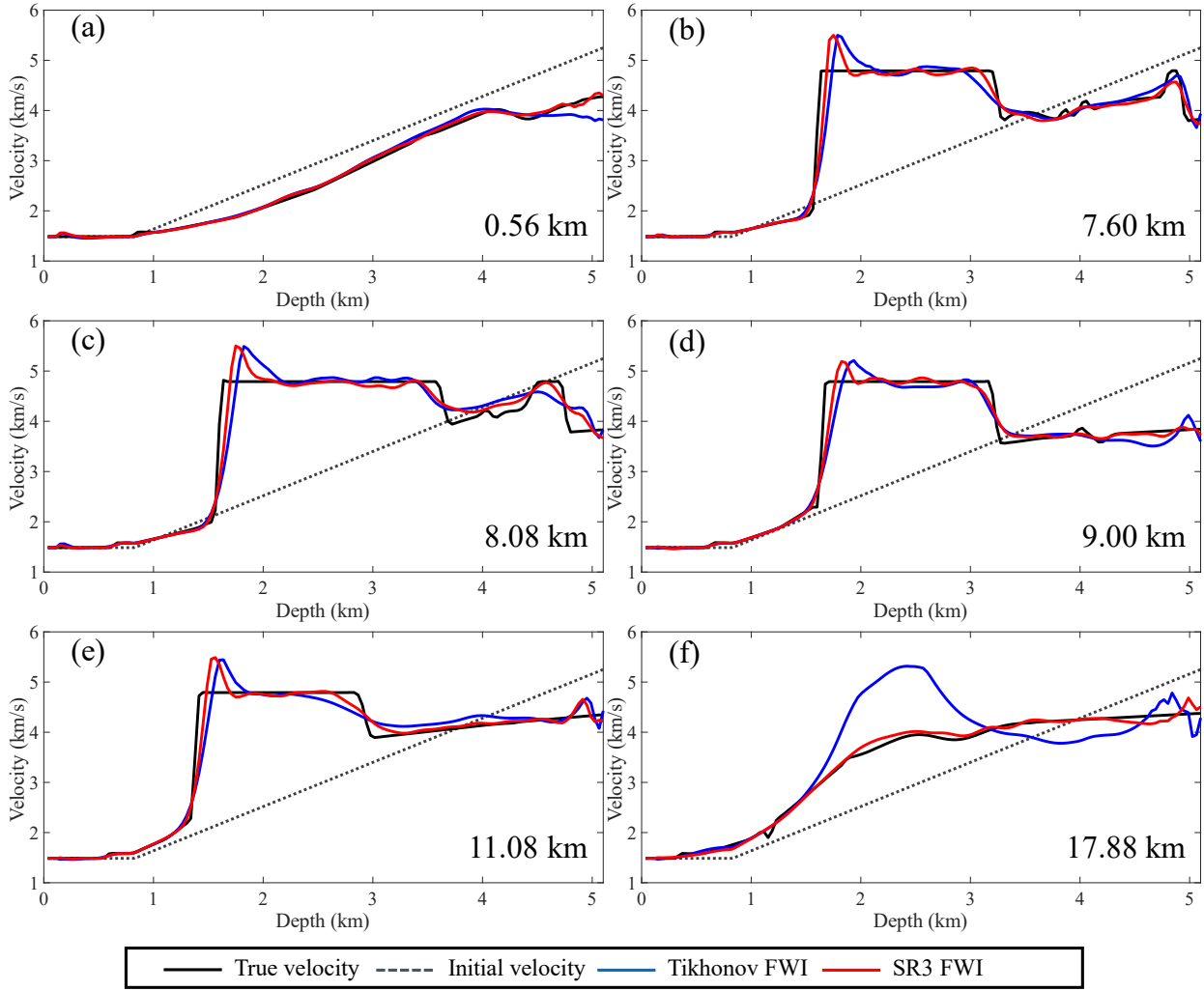

**Figure 14.** 2004 BP model, one-dimensional velocity models at different X-positions, (a) X = 0.56 km; (b) X = 7.60 km; (c) X = 8.08 km; (d) X = 9.00 km; (e) X = 11.08 km; (f) X = 17.88 km, the vertical comparison of the actual velocity model (solid black line), initial velocity model (grey dotted line), the Tikhonov regularisation FWI velocity model (solid blue line), and the SR3-based FWI velocity model (solid red line).

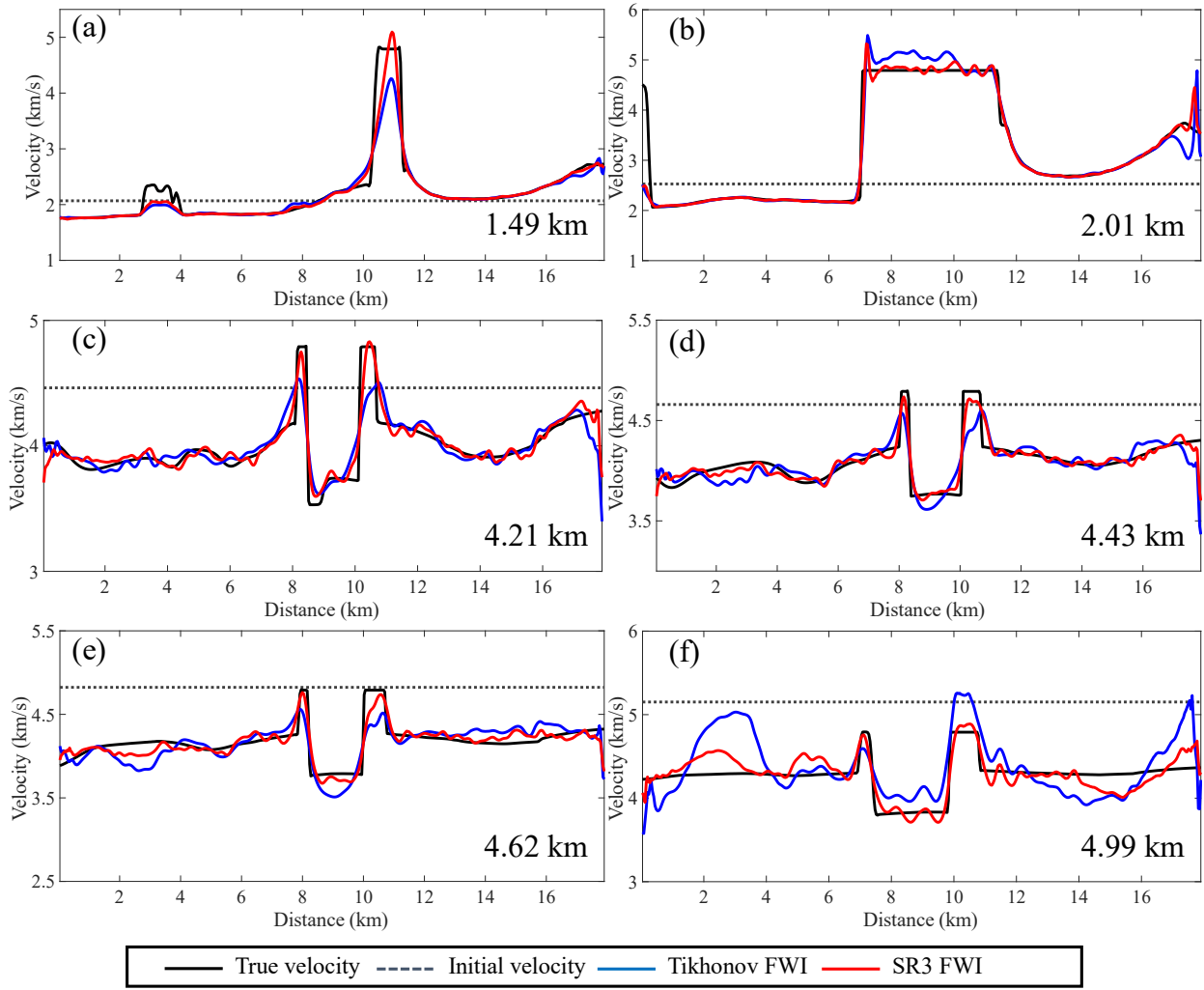

**Figure 15.** 2004 BP model, one-dimensional velocity models at different *Y*-positions, (a) *Y* = 1.49 km; (b) *Y* = 2.01 km; (c) *Y* = 4.21 km; (d) *Y* = 4.43 km; (e) *Y* = 4.62 km; (f) *Y* = 4.99 km, the horizontal comparison of the actual velocity model (solid black line), initial velocity model (grey dotted line), the Tikhonov regularisation FWI velocity model (solid blue line), and the SR3-based FWI velocity model (solid red line).

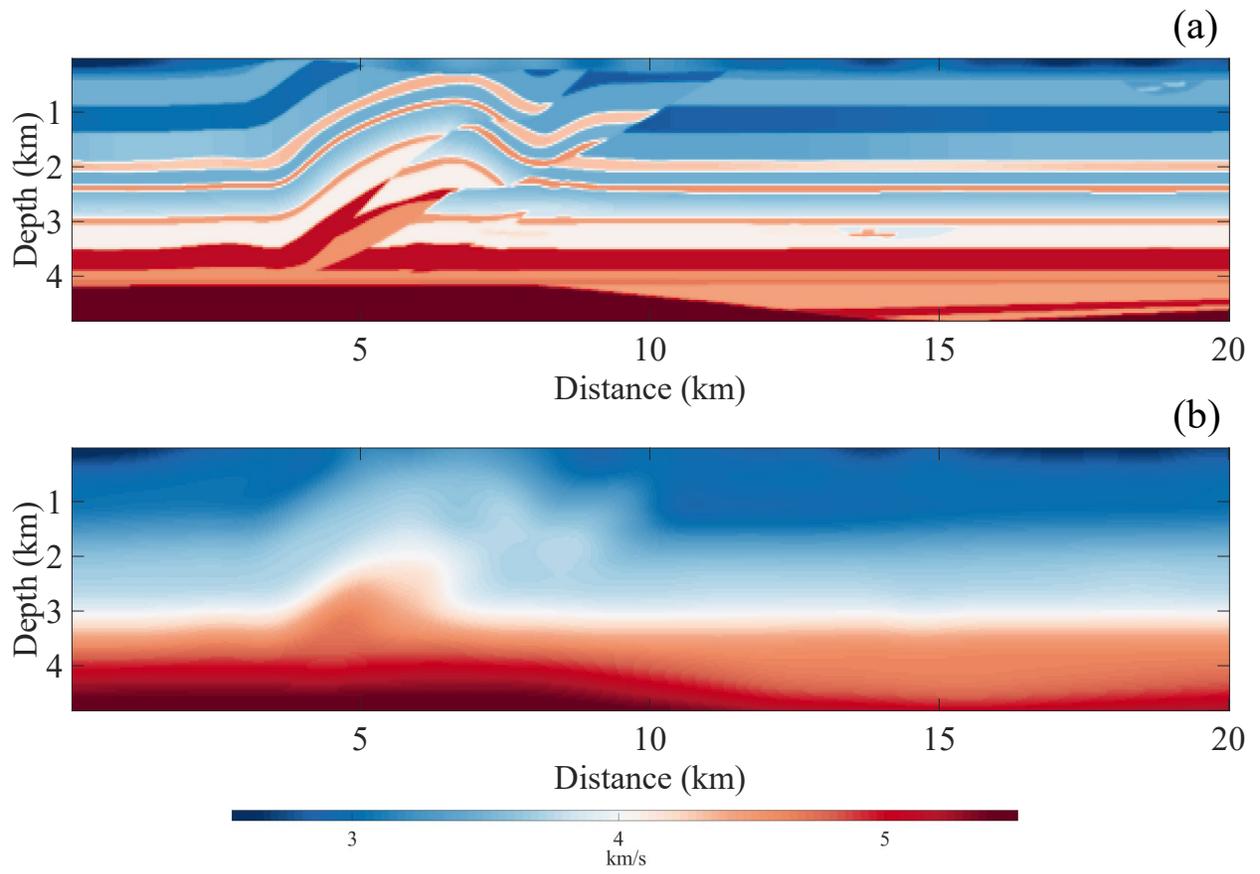

**Figure 16.** SEG/EAGE overthrust model, (a) true velocity model. (b) initial velocity model.

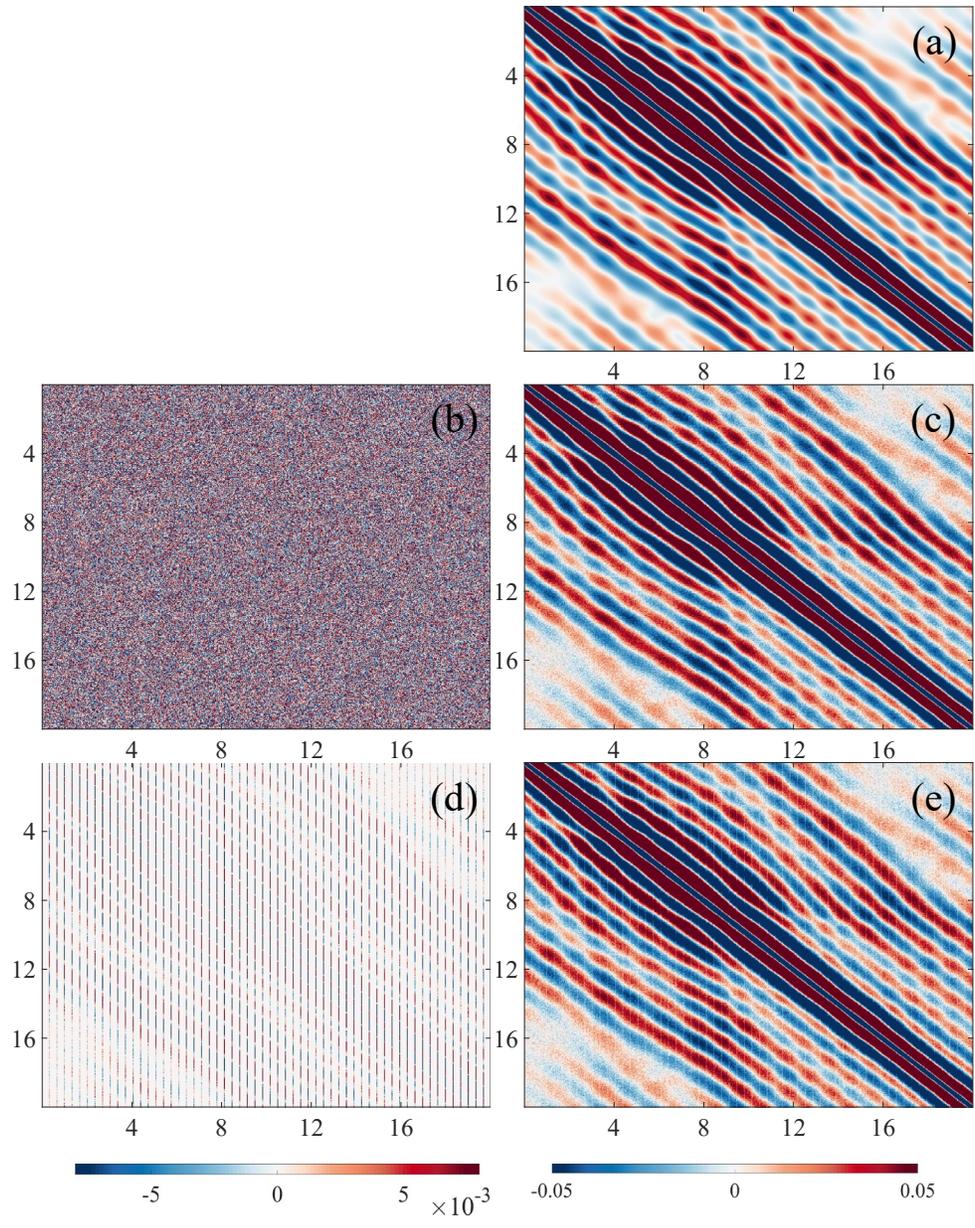

**Figure 17.** Source-receiver domain data set at 3 Hz of the SEG/EAGE overthrust model, the real part of the (a) clean data matrix, (b) 10 *dB* random noise, (c) wavefield matrix with 10 *dB* noise, (d) missing-trace matrix, (e) subsampled wavefield matrix with 10 *dB* noise and missing trace.

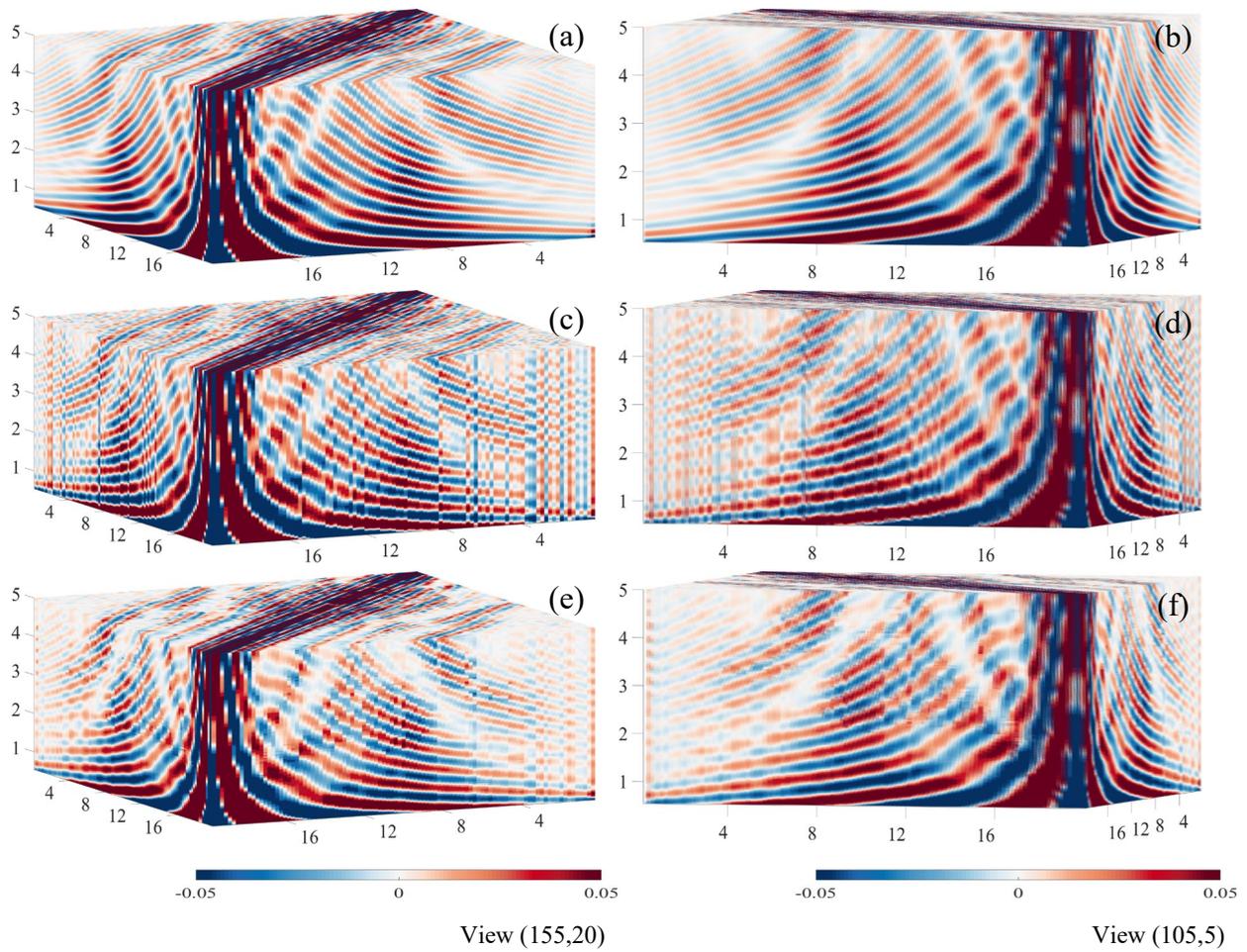

**Figure 18.** Three-dimensional low-frequency source-receiver wavefield from 0.5 Hz to 5 Hz of the SEG/EAGE overthrust model, (a) clean data matrix, (b) side view of (a); (c) subsampled wavefield matrix with missing-trace and 10 *dB* noise, (d) side view of (c); (e) wavefield matrix after SR3 optimization for (c), (f) side view of (e).

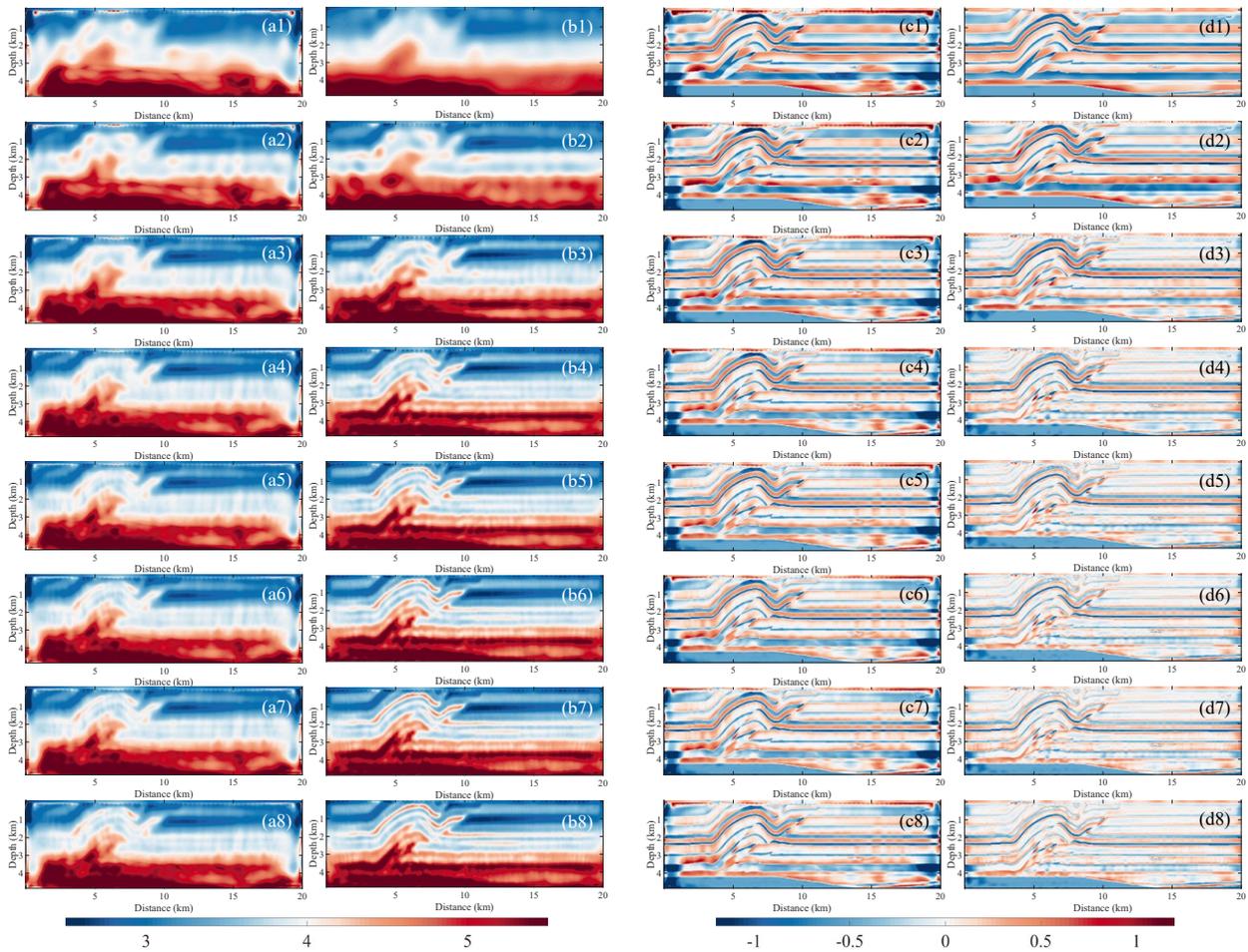

**Figure 19.** The SEG/EAGE overthrust model inversion results. (a1-a8) Tikhonov regularisation FWI inversion results in 2.15 Hz, 3.09 Hz, 5.35 Hz, 7.70 Hz, 9.24 Hz, 11.09 Hz, 13.31 Hz, and 15.97 Hz, respectively; (b1-b8) FWI based on SR3 algorithm inversion results in 2.15 Hz, 3.09 Hz, 5.35 Hz, 7.70 Hz, 9.24 Hz, 11.09 Hz, 13.31 Hz, and 15.97 Hz, respectively; (c1-c8) differences between the Tikhonov FWI and the true velocity model; (d1-d8) differences between the SR3 FWI and the true velocity model.

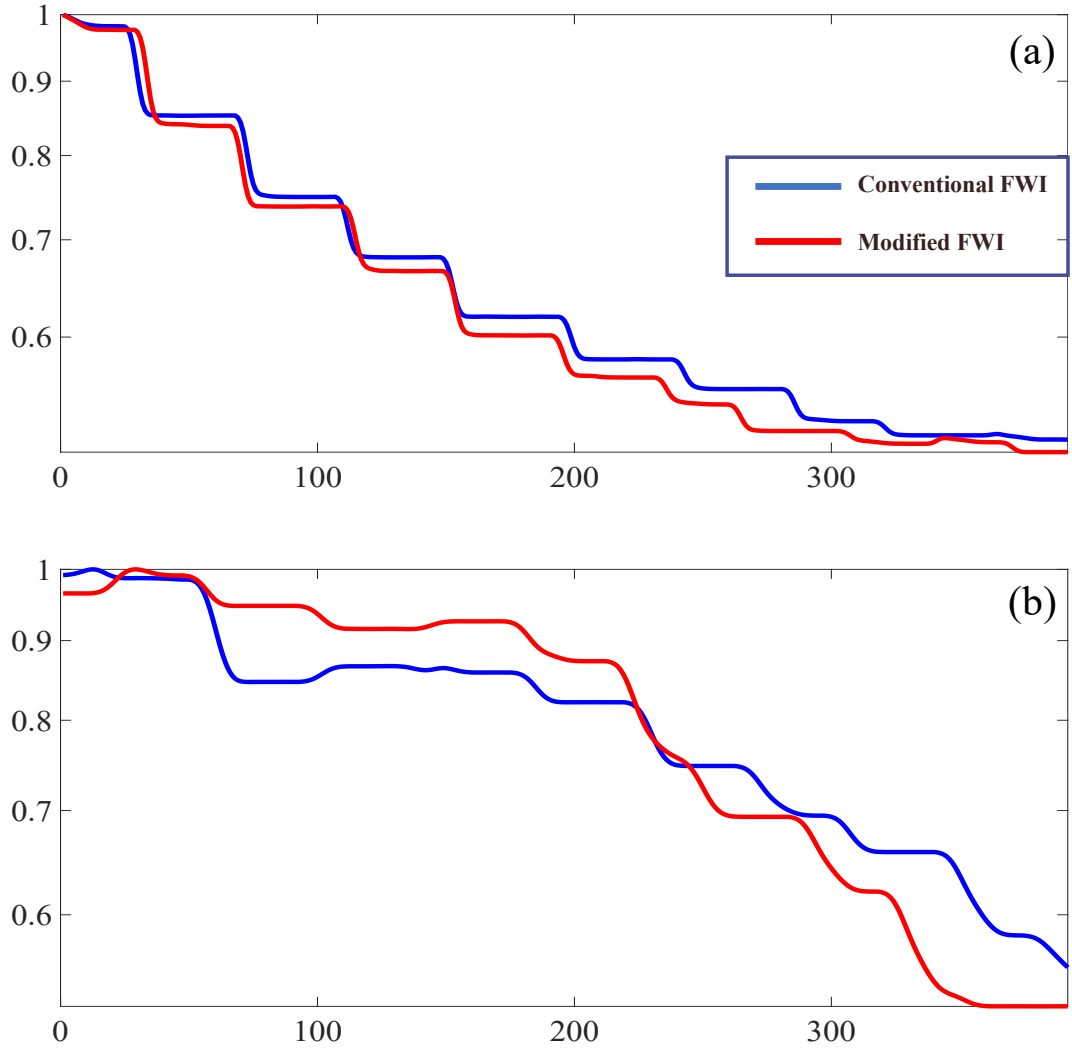

**Figure 20.** Comparison of SR3 algorithm-based FWI with conventional Tikhonov regularisation-based FWI for quantification, (a) normalised misfit error, (b) model error. The horizontal axis is the number of iterations, and the vertical axis is the error value.

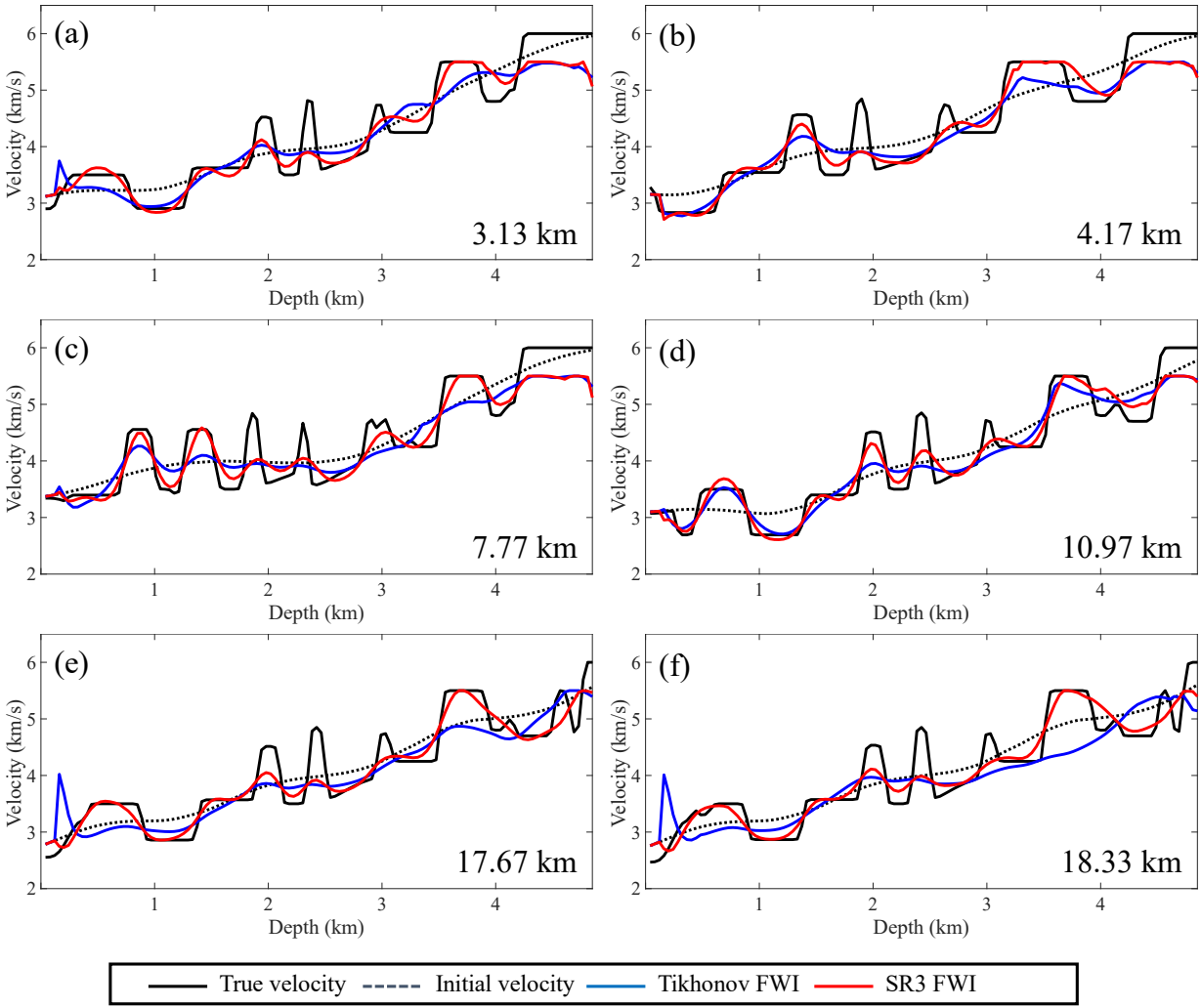

**Figure 21.** The SEG/EAGE overthrust model, 1-D velocity models at different *X*-positions, (a) *X* = 3.13 km; (b) *X* = 4.17 km; (c) *X* = 7.77 km; (d) *X* = 10.97 km; (e) *X* = 17.67 km; (f) *X* = 18.33 km. The vertical comparison of the actual velocity model (solid black line), initial velocity model (grey dotted line), the Tikhonov regularisation FWI velocity model (solid blue line), and the SR3-based FWI velocity model (solid red line).

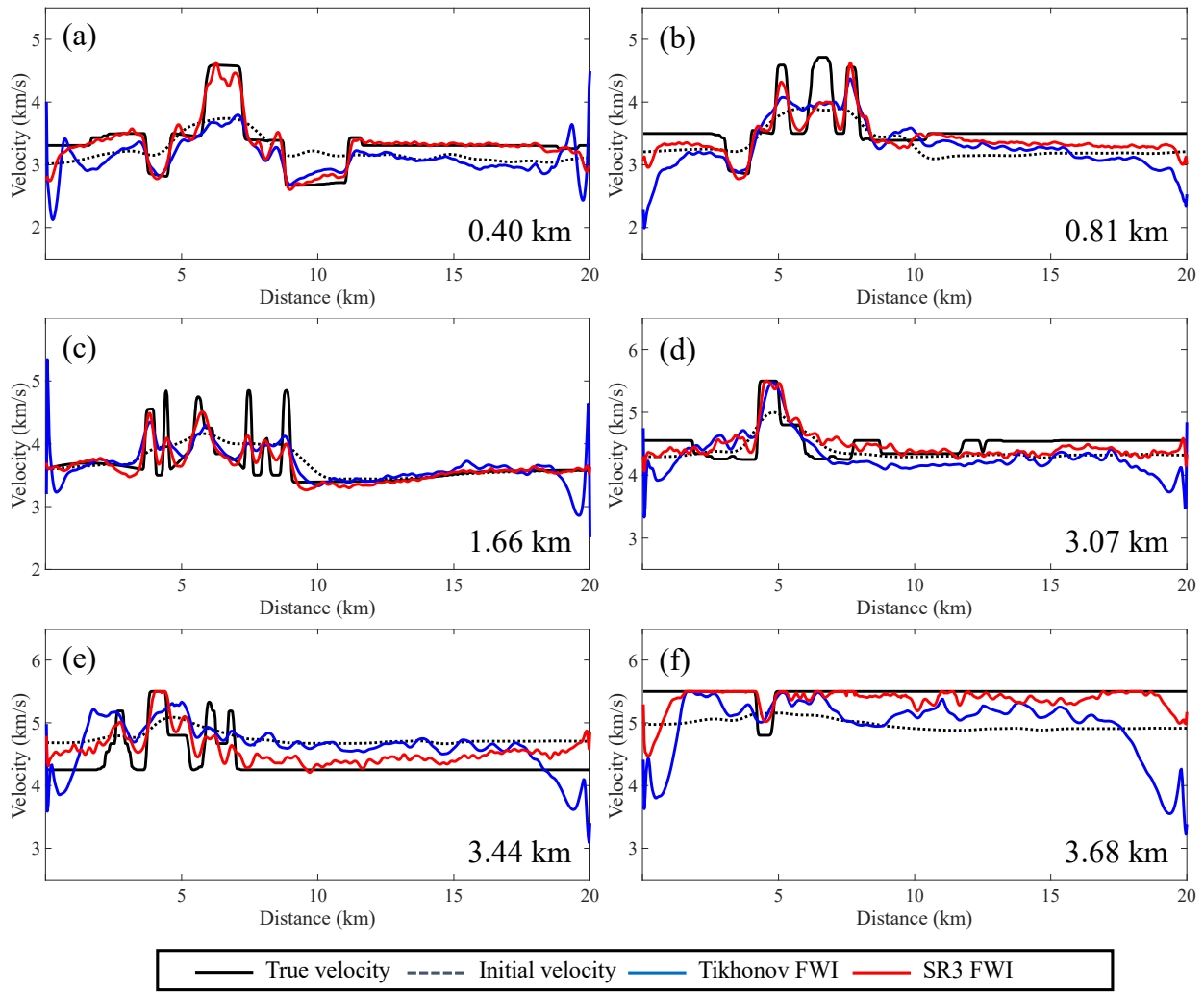

**Figure 22.** The SEG/EAGE overthrust model, 1-D velocity models at different *Y*-positions, (a) *Y* = 0.40 km; (b) *Y* = 0.81 km; (c) *Y* = 1.66 km; (d) *Y* = 3.07 km; (e) *Y* = 3.44 km; (f) *Y* = 3.68 km. The horizontal comparison of the actual velocity model (solid black line), initial velocity model (grey dotted line), the Tikhonov regularisation FWI velocity model (solid blue line), and the SR3-based FWI velocity model (solid red line).